\documentstyle[12pt,amssymb]{amsart}

\ifx\mathrm\undefined\let\mathrm\bf\fi
\ifx\mathbf\undefined\let\mathbf\bf\fi
\ifx\mathfrak\undefined\let\mathfrak\frak\fi
\ifx\mathcal\undefined\let\mathcal\cal\fi
\ifx\mathbb\undefined\let\mathbb\Bbb\fi
\ifx\emph\undefined\let\emph\it\fi

\newcommand{\Z}{{\mathbb Z}}

\newcommand{\C}{{\mathbb C}}

\newcommand{\Pee}{{\mathbb P}}
\newcommand{\Ref}[1]{{(\ref{#1})}}
\newcommand{\bean}{\begin{eqnarray}}
\newcommand{\eean}{\end{eqnarray}}
\newcommand{\be}{\begin{displaymath}}
\newcommand{\ee}{\end{displaymath}}
\newcommand{\bea}{\begin{eqnarray*}}
\newcommand{\eea}{\end{eqnarray*}}
\newcommand{\g}{{{\mathfrak g}\,}}

\newcommand{\Id}{{\mathrm{Id}}}
\newcommand{\ad}{{\mathrm{ad}}}
\newcommand{\res}{{\mathrm{res}}}

\newcommand{\noi}{\noindent}
\newcommand{\into}{\hookrightarrow}
\newcommand{\dds}{\frac d{ds}\bigg|_{s=0}}
\newcommand{\vs}{\vspace{1.5\baselineskip}}
\newenvironment{prf}{\noindent{\it Proof\/}:}{$\;\Box$
\par\vs}

\newtheorem%
{thm}{Theorem}[section]
\newtheorem%
{proposition}[thm]{Proposition}
\newtheorem%
{lemma}[thm]{Lemma}
\newtheorem%
{lemmadef}[thm]{Lemma-Definition}
\newtheorem%
{corollary}[thm]{Corollary}
\newtheorem%
{conjecture}[thm]{Conjecture}
\newenvironment{definition}
{\noindent{\bf Definition\/}:}{\par\vs}
\newcommand{\ddz}{{\frac{d}{dz}}}
\newcommand{\End}{{\mathrm{End}}}

\newcommand{\tr}{{\mathrm{tr}}}
\newcommand{\Ad}{{\mathrm{Ad}}}
\newcommand{\Diff}{{\mathrm{Diff}_0(\Sigma,S,(v_j))}}
\title
{The KZB equations on Riemann surfaces}
\author{Giovanni Felder}
\date{}
\begin{document}
\maketitle
\centerline{Departement Mathematik, ETH-Zentrum,}
\centerline{CH-8092 Z\"urich, Switzerland}

\begin{abstract} In this paper, based on the author's lectures at the
1995 les Houches Summer school, explicit expressions for the
Friedan--Shenker connection on the vector bundle of WZW conformal
blocks on the moduli space of curves with tangent vectors at $n$
marked points are given. The covariant derivatives are expressed in
terms of ``dynamical $r$-matrices'', a notion borrowed from integrable
systems. The case of marked points moving on a fixed Riemann surface
is studied more closely. We prove a universal form of the (projective)
flatness of the connection: the covariant derivatives commute as
differential operators with coefficients in the universal enveloping
algebra -- not just when acting on conformal blocks.
\end{abstract}

\section{Introduction}

The Knizhnik--Zamolodchikov--Bernard, or KZB, equations are remarkable
systems of differential equations which generalize the Gauss
hypergeometric equation. There is such a system of differential
equations for each simple complex Lie group $G$, a set of
representations $V_1,\dots, V_n$ of $G$, an integer  $k$
and a non-negative integer $h$, with some restrictions. The
equations are the equations of horizontality for 
a local section $u$ of a vector bundle with connection,
 the vector bundle of conformal 
blocks, over the
moduli space $\hat M_{h,n}$ of Riemann surfaces of genus $h$ with
tangent vectors at $n$ marked points.

These equations have been studied mostly in the case of genus
zero (the Knizhnik--Zamolodchikov equations), 
with a surprising range of applications, from number 
theory to quantum integrable systems to three-dimensional topology.

The genus one equations have also been studied in some detail.
In particular, integral representations of solutions of hypergeometric
type are known now both in genus zero and in genus one, see
\cite{SV}, \cite{FV1}.

For  higher genus Riemann surfaces, the equations have not been
studied in detail. One does know of the existence of the 
connection, and that it extends to a compactification of moduli
space, leading to factorization theorems, but not much more is known.
The object of this paper is a description in concrete terms of
the KZB equations in genus $\geq2$. This is done in the original
papers by Bernard \cite{B1,B2} for a particular parametrization of the
moduli space of $G$-bundles based on a Schottky representation
of the Riemann surfaces. In the case $n=0$ a rather explicit
and general description is given by Hitchin in \cite{H1}.
 Here we consider general coordinates
on the moduli spaces, and relate the formulae to some familiar
objects ($r$-matrices, $\ell$-operators) of the theory of
classical integrable systems. The setting is not quite the
same as in integrable systems, so these objects come with 
a twist here and there.

The origin of the KZB equations is in the WZW model of conformal
field theory \cite{W,KZ}. The central idea, formulated by Friedan and Shenker
\cite{FS}, is that the connection is given by the energy-momentum
tensor of the quantum field theory. The approach we take
will be close to the original one of Friedan and Shenker, but
we translate the notions to a more mathematical setting.

Here is an outline of the construction.
Associated to the data $V_1$,\dots, $V_n$ and $k$ large enough,
we have a vector bundle over a suitable
compactification of the moduli space of stable $G$-bundles on
a Riemann surface $\Sigma$ with marked points $p_1,\dots,p_n$.

A conformal block is a global holomorphic section of this
vector bundle. The spaces of conformal blocks form a vector
bundle over the moduli space $\hat M_{h,n}$.
One then defines a ``current insertion'' $J_x(p)$
associated with a point $p$ distinct from the marked points.
It is a first order differential operator on conformal blocks
corresponding to the vector field on the space of $G$-bundles
defined roughly by the following local surgery procedure. If
we have a $G$-bundle $P$, with given trivialization around $p$,
and local coordinate $t$, the vector field points in the direction
of the $G$-bundle obtained by taking out the fibers of $P$ over
a neighborhood of $p$ and gluing them back with transition function
$\exp(\epsilon x/(t-t(p)))$. The energy-momentum tensor $T(p)$ is
essentially a suitable quadratic expression in the currents. 

These operators depend on choices and do not map conformal blocks to
conformal blocks. However, if $u$ is a section of the vector bundle of
conformal blocks then the covariant derivative $\nabla_\zeta u=
\partial_\zeta u+\langle T,\zeta\rangle u$ is again
a conformal block. Here $\zeta\in H^1(\Sigma,K^2\otimes(-2\sum p_i))$
is a tangent vector that pairs with $T$, which depends on ``flat''
coordinates
as a quadratic differential. See below for a more precise statement.

To make this explicit we choose to work in the double coset
representation of the moduli space of $G$-bundle.  We consider the
case $G=SL(N,\C)$ for simplicity of exposition.  If $U_S$ is a
neighborhood of $S=\{p_1,\dots,p_n\}$, consisting of little disjoint
open disks arount the points, isomorphism classes of $G$-bundles are
in one-to-one correspondence with double cosets in 
${\mathcal{M}}_G=G(U_S)\backslash G(U_S^\times)/ G(\Sigma-S)$. 
Here $G(X)$ denotes
the infinite-dimensional complex Lie group of holomorphic maps
from the complex manifold $X$ to $G$, and $U_S^\times=U_S-S$. 
The group $G(U^\times_S)$ has a central extension $\hat G(U_S^\times)$
by $\C^\times$ which splits over the two subgroups, and is 
a principal $\C^\times$-bundle  over $G(U_S^\times)$. The space
of conformal blocks, cf.~\cite{S}, is the space of holomorphic functions
$u(\hat g)$ on $G(U_S^\times)$ with values in $V=V_1\otimes\cdots
\otimes V_n$ such that $u(\hat gz)=z^ku(\hat g)$
for all $z\in\C^\times$ and $u(b\hat g n)=bu(\hat g)$ for all
$b\in G(U_S)$, $n\in G(\Sigma-S)$. The group $G(U_S)$ acts on the
values by $bu=b(p_1)\otimes\cdots\otimes b(p_n)u$.

We will not be concerned here with the proper definitions of 
holomorphy in this infinite dimensional setting, and refer the
interested reader to the papers \cite{BL,Fa,KNR}.
The main result is that, with proper definitions, the space of
conformal blocks is finite dimensional, and if we have any
holomorphic family of Riemann surfaces with  marked points,
the spaces of conformal blocks form a holomorphic vector bundle
over the parameter space.

The form of the connection is then, in terms of coordinates
$\lambda_1,\dots,\lambda_n$ on the moduli space of $G$-bundles,
and local trivialization
\bea
\nabla_\zeta u(\tau,\lambda)=
\partial_\zeta u(\tau,\lambda)+\frac 1{2\pi i}
\oint_\gamma A(z)\zeta(z)dz\,u(\tau,\lambda),
\eea
for a second order differential operator $A(z)$ depending on
coordinates like a quadratic differential (given a 
``flat structure'' on $\Sigma$). Here $\gamma$ is
a contour winding around every point in $S$ once, and a tangent
vector $\zeta$ is represented by a holomorphic vector
field on a pointed neighborhood $U_S^\times$ of $S$.  The differential
operator $A(z)$ is given by an expression of the form
\be
A(z)=\frac 1{2(k+h^\vee)}(\tr^{(0)}\hat \ell(z)^2
+\tr^{(0)} q_1(z,\lambda)^{(0)}\hat\ell(z,\tau)+kq_2(z,\lambda))
\ee
The trace is the ``trace over the auxiliary space'' from integrable
models: the object $\hat \ell(z)$ belongs to $\g\otimes\End(V)\otimes
{\mathcal D}$, where ${\mathcal D}$ denotes the space of differential
operators in the variables $\lambda_j$, and $\tr^{(0)}$ is the
trace over the first factor in the tensor product.
The $\ell$-operator $\hat\ell(z)$ is built out of the $r$-matrix
by the formula
\be
\hat\ell(z)=\sum_{i=1}^m\omega_i(z,\lambda)^{(0)}\partial_{\lambda_i}
+\sum_{j=1}^nr(z,p_j,\lambda)^{(0j)}+kq(z,\lambda)^{(0)}.
\ee
In this expression $\omega_i(z,\lambda)dz$ is a basis of 
$H^0(\Sigma,\Ad(P_\lambda)\otimes K)$ Serre dual to the
basis $\partial/\partial\lambda_j$ of the tangent space
to the moduli space of $G$-bundle at the class of the $G$-bundle
$P_\lambda$ parametrized by $\lambda$. The ``dynamical $r$-matrix''
$r(z,t,\lambda)\in\g\otimes\g$  is the
building block of all the other quantities. It is defined in
\ref{drm}. For instance, $q_1$ is the Lie bracket of the
constant term in the Laurent expansion of $r$ at its pole
on the diagonal, and $q_2$ is obtained by applying the invariant
bilinear form on the first order Laurent coefficient, see
below.

\section{Conformal blocks on  Riemann surfaces}
\subsection{Kac--Moody groups}\label{sec2} Let $G$ be a simply connected
complex simple Lie group, with Lie algebra
$\g$. For simplicity of exposition, we will
assume that $G={\mathrm{SL}}(N,\C)$, the group
of $N$ by $N$ matrices with complex entries and unit determinant.

For any complex manifold $U$, we let $G(U)$ be the group of holomorphic
maps from $U$ to $G$ with pointwise multiplication. Is is  an infinite
dimensional complex Lie group. Its Lie algebra
is the Lie algebra $\g(U)$ of holomorphic maps $U\to\g$.

Let $U$ be an open disk containing the origin in the complex plane,
and $U^\times=U-\{0\}$. The loop algebra
$\g(U^\times)$ has a universal central extension $\hat\g(U^\times)$
(the affine Kac-Moody algebra)
corresponding to the two-cocycle 
\begin{equation}\label{Coc}
c(x,y)=\res_{t=0}\tr(x'(t)y(t))dt
\end{equation}
On the group level, see \cite{PS}, we have a corresponding central extension 
\be
1\to\C^\times\to\hat G(U^\times)\stackrel\pi\to G(U)
\to 1
\ee 
which we now
describe. For any smooth map $g:\Sigma\to G$ on a compact Riemann
surface $\Sigma$ with values in $G$, the WZW action
\cite{W} is ($\partial=dz\partial/\partial z$, $\bar\partial=
\partial/\partial\bar z$ for any choice of local coordinate $z$)
\be 
W(g)=-\frac i{4\pi}\int_{\Sigma}\tr (g^{-1}\partial g
\,g^{-1}\bar\partial g)+\frac i{12\pi}\int_{B}\tr((g^{-1}dg)^3) \in
\C/2\pi i\,\Z.
 \ee 
In the second (Wess--Zumino) term, the Riemann surface
$\Sigma$ is viewed as the boundary of a three-dimensional
manifold $B$ and the map $g$ is extended to a map from
$B$ to $G$. It can be shown that this is possible and that the
integrals obtained by choosing different $B$'s or different
extensions differ by integer
multiples of $2\pi i$. This follows from the fact that the difference
of integrals corresponding to different choices is the integral
over a closed three-manifold of the pull-back of the three-form
$\frac i{12\pi}\tr((g^{-1}dg)^3)$ on $G$, which is in (the image
of) $H^2(G,2\pi i\,\Z)$. In the case of general simply connected
$G$, one replaces
$\tr(xy)$ by $(x,y)$, an invariant symmetric bilinear form,
and $i(12\pi)^{-1}\tr((g^{-1}dg)^3)$ by $i(12\pi)^{-1}
([g^{-1}dg,g^{-1}dg],g^{-1}dg)$,
which is still a $2\pi i\times$integral three-form on $G$, provided 
the bilinear form is normalized in such a way that long
roots have length squared two. In the non-simply connected
case, things are slightly more tricky, see \cite{FGK}.

Let, for small $\epsilon>0$, $U_\epsilon=\{z\in U|\,|z|>
\epsilon\}$ and let us first define the
 group 
$\hat G(U_\epsilon)$ consisting of equivalence classes of 
pairs $(g,z)$ where $g$ is a smooth map from $U$ to $G$, holomorphic
on $U_\epsilon$ and
$z$ is a non-zero complex number. Two pairs $(g_1,z_1)$,
$(g_2,z_2)$ are equivalent if $g_2=g_1h$ for some smooth map $h:U\to G$
with  $h|_{U_\epsilon}=1$ and
\be
z_2=z_1e^{W(h)+\frac1{2\pi i}\int_U\omega(g_1,h)},
\qquad
\omega(g,h)=
\tr(g^{-1}\partial g\bar\partial h h^{-1}).
\ee
Here $W(h)$ is defined by extending $h$ to a smooth map from the Riemann
sphere $\Pee^1=\C\cup\{\infty\}$ to $G$ such that  $h(z)=1$ if $|z|>\epsilon$.
Let us denote by $[(g,z)]$ the equivalence class of $(g,z)$.

The product in $\hat G(U_\epsilon)$ is defined by the rule
\be
[(g_1,z_1)][(g_2,z_2)]=[(g_1g_2,z_1z_2\,e^{\frac1{2\pi i}
\int_U\omega(g_1,g_2)})],
\ee
and the inverse of $[(g,z)]$ is $[(g^{-1},z^{-1}\exp(-\frac1{2\pi i}
\omega(g,g^{-1})))]$.
The WZW action obeys the product formula
\be
W(gh)=W(g)+W(h)+\frac{1}{2\pi i}\int_\Sigma\omega(g,h)
\ee
for smooth maps $g, h:\Sigma\to G$. If $\Sigma=\Pee^1$, this
formula can be used to check that the above relation is indeed an
equivalence relation, and that the product is well defined on
equivalence classes.
Then, by construction, we have a central extension
\be
1\to\C^\times\to\hat G(U_\epsilon^\times)\to G(U_\epsilon^\times)\to 1.
\ee
The nontrivial maps are $z\mapsto(1,z)$ and $(g,z)\mapsto g$.
We may define $\hat G(U^\times)$ as the inverse image of $G(U^\times)
\subset G(U_\epsilon)$. This is independent of the choice of $\epsilon$ up to
isomorphism:
\begin{proposition}
\ \ \

\begin{enumerate}
\item[(i)] 
If $\epsilon_1<\epsilon_2$ are small,  the restriction homomorphism
 $G(U_{\epsilon_2})\into G(U_{\epsilon_1})$
lifts uniquely
 to an embedding $j:\hat G(U_{\epsilon_2})\into
\hat G(U_{\epsilon_1})$ preserving the central subgroup $\C^\times$. On
the inverse image of $G(U^\times)$, $j$ is an isomorphism.
\item[(ii)]
If $U_1\subset U_2$ then the restriction homomorphism
 $G(U_2)\into G(U_1)$
lifts uniquely
 to an embedding $j:\hat G(U_2)\into
\hat G(U_1)$ preserving the central subgroup $\C^\times$.
\end{enumerate}
\end{proposition}
The uniqueness follows from the fact that two lifts
differ by a homomorphism from $G(X)$ to $\C^\times$, for appropriate $X$.
Since $\C^\times$ is Abelian, any such homomorphism vanishes
on the commutator subgroup. But $G(X)$ is equal to its commutator
subgroup, see \cite{PS}, Chapter 3.

Thus an element of $\hat G(U^\times)$ is an equivalence class
of pairs $(g,z)$ such that $g:U\to G$ is smooth and coincides
with an an element of $G(U^\times)$ except on some small neighborhood
of the origin.

By the property (ii), we can also pass to the inverse limit 
$\hat G=\lim_{U\ni 0}\hat G(U^\times)$.

The central extension $\hat G(U^\times)$ splits over some subgroups
of $G(U^\times)$:
\begin{thm}\label{subgroups}
\ \

\begin{enumerate}
\item[(i)] The map $g\mapsto [(g,1)]$ is an injective homomorphism $j$
of $G(U)$ into $\hat G(U^\times)$ with $\pi\circ j=\Id$.
\item[(ii)] Let $U$ be embedded into a compact Riemann surface
$\Sigma$, so that $0$ is mapped to a point $p\in\Sigma$. Consider
$G(\Sigma-\{p\})$ as the  subgroup of $G(U^\times)$ of the maps
extending to $\Sigma-\{p\}$. Then there is an injective
homomorphism $j:G(\Sigma-\{p\})\into\hat G(U^\times)$ with
$\pi\circ j={\mathrm{Id}}$. It is given explicitly by the formula
\be
g\mapsto [(\tilde g,e^{W(\tilde g)})]
\ee
for any smooth map $\tilde g:\Sigma\to G$ which coincides with
$g$ on the complement of a small neighborhood of $p$.
\end{enumerate}
\end{thm}

We conclude this section by describing the 
Lie algebra ${\mathrm{Lie}}(\hat G(U^\times))$ of 
$\hat G(U^\times)$. First of all, the Lie algebra $\g(U^\times)$
can be canonically embedded as a vector space into 
${\mathrm{Lie}}(\hat G(U^\times))$:
if $x\in\g(U^\times)$ let $x_{\mathrm{reg}}$, a regularization
of $x$, be any smooth map $U\to\g$
which coincides  with $x$ except possibly on some small neighborhood
of the origin. Then we define the right-invariant vector field
$D_x$ acting on holomorphic functions on $\hat G(U^\times)$ by
\be
D_xf(g)=\frac d{ds}\bigg|_{s=0}f([(\exp(-sx_{\mathrm{reg}}),1)]g)
\ee
This definition is independent of the choice of regularization
$x_{\mathrm{reg}}$: the difference of the vector fields corresponding
to different regularizations is $D_yf(g)=\frac d{ds}|_{s=0}f([(h_s,1)]g)$
where $h_s=\exp sy$, and $y(z)=0$ for $|z|>\epsilon$. But 
$(h_s,1)\sim(1,\exp(-W(h_s)))$, so $D_yf=0$ since 
$W(h_s)=O(s^2)$.

The Lie algebra ${\mathrm{Lie}}(\hat G(U^\times))$ is thus spanned
by $D_x$, $x\in\g(U^\times)$ and the generator $K$ of the center:
$Kf(g)=\frac {d}{ds}|_{s=0}f([(1,e^s)]g)$. 

\begin{proposition}\label{propLie}
The Lie brackets on ${\mathrm{Lie}}(\hat G(U^\times))$ are
given by the formulae (see \Ref{Coc})
\bea
[D_x,K]&=&0,
\\
{}[D_x,D_y]
&=&
D_{[x,y]}+c(x,y)K, \qquad x,y\in\g(U^\times).
\eea
Thus ${\mathrm{Lie}}(\hat G(U^\times))$ 
 is isomorphic to the affine algebra $\hat g(U^\times)$.
\end{proposition}
\begin{prf}
The first bracket follows from the fact that $K$ is in the Lie algebra of
a central subgroup.
Let $g_t=\exp(-t x_{\mathrm{reg}})$ and $h_s=\exp(-s y_{\mathrm{reg}})$,
for some choices of regularization. By definition,
\be
D_xD_yf(g)=\frac d{dt}\bigg|_{t=0}\frac d{ds}\bigg|_{s=0}
f([(h_s,1)][(g_t,1)]g).
\ee
Neglecting terms of order higher than one in $t$ or $s$, which
do not contribute to the derivative at zero, we have
\bea
[(h_s,1)][(g_t,1)]&=&[(h_sg_t,e^{\frac1{2\pi i}\int_U\omega(h_s,g_t)})]
\\
 &=&[(g_th_s\exp(-st[x,y]),e^{\frac1{2\pi i}\int_U\omega(h_s,g_t)})]
\\
&=&[(g_t,1)][(h_s,1)]
\\
& &[(\exp(-st[x,y]),
e^{\frac1{2\pi i}\int_U(\omega(h_s,g_t)
-\omega(g_t,h_s))})].
\eea
Therefore,
\bea
D_xD_yf&=&D_yD_xf+D_{[x,y]}f\\
 & &+\frac d{dt}\bigg|_{t=0}\frac d{ds}\bigg|_{s=0}
\frac1{2\pi i}\int_U(\omega(h_s,g_t)
-\omega(g_t,h_s))Kf.
\eea
The coefficient of $Kf$ is
\be\frac1{2\pi i}\int_U\tr(dy_{\mathrm{reg}} dx_{\mathrm{reg}})=
\frac1{2\pi i}\oint\tr(y\,dx),
\ee
by Stokes' theorem. The contour of integration is chosen to lie
in the region where $x_{\mathrm{reg}}$, $y_{\mathrm{reg}}$ 
coincide with
$x$ and $y$. This integral is the residue at zero, which is what
had to be proven.
\end{prf}

\subsection{Principal $G$-bundles}\label{sec3}
Let $\Sigma$ be a compact Riemann surface and $S=\{p_1,\dots,p_n\}$
be a finite set of $n\geq 1$ points on $\Sigma$.
Let $U_S$ be a neighborhood of $S$ consisting of $n$ embedded disjoint
disks around the points $p_i$, and $U^\times_S=U_S-S$. Consider
the group $G(U_S^\times)$ of holomorphic maps $U_S^\times\to G$.
The group multiplication is defined pointwise. The group $G(U^\times_S)$
has the subgroups $G(U_S)$ of holomorphic maps from $U_S$ to $G$,
and $G(\Sigma-S)$ of holomorphic maps from $\Sigma-S$ to $G$.

For each element $g$ of $G(U_S^\times)$ 
we have a principal $G$-bundle $P_g$
on $\Sigma$: it is defined as the quotient space
$P_g$ of the disjoint union $U_S\times G\, \sqcup\, (\Sigma-S)\times G$, 
by the
relation $U_S\times G \ni(z,h)\sim (z,g(z)^{-1}h)\in (\Sigma-S)\times G$,
for all $z\in U_S^\times, h\in G$. The projection $P_g\to\Sigma$ is
the projection onto the first factor, and the action of $G$ is the
right action on the second factor. Two $G$-bundles $P_{g}$, $P_{h}$ 
are isomorphic if and only if there exist elements $b\in G(U_S)$
and $n\in G(\Sigma-S)$ such that 
$h(z)=b(z)g(z)n(z)$ for all $z\in U_S^\times$. Moreover it can be shown
that every holomorphic $G$-bundle on $\Sigma$ is isomorphic to some
$P_g$. For $SL(N,\C)$ this is the content of Grauert's theorem,
see \cite{Gr}, Chapter 8.
Thus equivalence classes of $G$-bundles are in one-to-one
correspondence with double cosets in $M_S=G(U_S)\backslash G(U_S^\times)/G(\Sigma-S)$.

\subsection{Conformal blocks}\label{scb}
The group $G(U_S^\times)$ has a central extension $\hat G(U_S^\times)$
by $\C^\times$, which is a principal $\C^\times$-bundle over $G(U_S^\times)$.
Let $U_S$ be the union of disks $U_i$ centered at $p_i$, and let
$U_i^\times=U_i-\{p_i\}$. Then
we have the direct product $\tilde G(U_S^\times)=\prod_{i=1}^nG(U_i^\times)$
of Kac-Moody groups, with its central subgroup $Z=(\C^\times)^n$.
By definition $\hat G(U_S^\times)$ is the quotient of
$\tilde G(U_S^\times)$ by the central subgroup 
\be
Z_0=\{(z_1,\dots,z_n)\in Z\,|\, z_1\cdots z_n=1\}.
\ee
Alternatively, we may proceed as in Section \ref{sec2} 
and define $\hat G(U_S^\times)$
directly to be the group of equivalence classes of pairs
$[(g,z)]$, where $g$ is a smooth map from $U_S$ to $G$ which coincides
with a map in $G(U_S^\times)$ except on some small neighborhood of $S$.
The equivalence relation is 
\be
(gh,ze^{W(h)+\frac1{2\pi i}\int_{U_S}
\omega(g,h)})\sim (g,z)
\ee for all
smooth maps $h:\Sigma\to G$, such that $h(q)=1$ outside a small neighborhood
of $S$.

The Lie algebra of $\hat G(U_S^\times)$ is spanned by vector fields
$D_x$, $x\in\g(U_S^\times)$ and the generator $K$ of the center.
The Lie brackets are as in Proposition \ref{propLie}, but the
cocycle is now
\be
c(x,y)=\sum_{j=1}^n\res_{p_j}\tr( dx\,y).
\ee
As before, the subgroups $G(U_S)$, $G(\Sigma-S)$ lift to subgroups of $\hat
G(U_S^\times)$ which we denote be the same letters.  

Let $V=V_1\otimes\cdots\otimes V_n$ be a tensor product of finite-%
dimensional irreducible representations of $G$. Let $G(U_S)$ act on
$V$ by $gv=g(p_1)\otimes\cdots\otimes g(p_n)v$.

\begin{definition}
Let $k$ be a non-negative integer.
The space of conformal blocks $E_{V,k}(\Sigma,S)$ of level $k$ is the
vector space of holomorphic functions $u:\hat G(U_S^\times)\to V$ such
that 
\be
u(bgn)=bu(g), \qquad u(gw)=w^ku(g),
\ee
for all $b\in G(U_S)$, $g\in\hat G(U_S^\times)$, $n\in G(\Sigma-S)$
and $w\in\C^\times$.
\end{definition}

This definition is in fact independent of the choice of the union of
disjoint disks $U_S$, in the sense that the restriction maps from the space of
conformal blocks defined using $U_S$ to the space defined using
$U_S'$ with $U_S'\supset U_S$ are isomorphisms.

Infinitesimally, the conditions on a conformal block $u$ are
as follows. Let $\gamma$ by a contour winding each of the points
$p_j$ once.

\begin{lemma}\label{l4}
Let $u\in E_{V,k}(\Sigma,S)$, and let $\hat g\in \hat G(U^\times)$ 
with $g=\pi(\hat g)\in G(U^\times_S)$. If $y\in\g(U^\times)$ then
\be
D_yu(\hat g)=-\sum y(p_j)^{(j)}u(\hat g),
\ee
 and if $\Ad(g^{-1})y$ extends
to a map in $\g(\Sigma-S)$, then
\be
D_yu(\hat g)+\frac k{2\pi i}\oint_\gamma\tr(dg\, g^{-1}y)u(\hat g)=0
\ee
Also, $Ku=ku$.
\end{lemma}

The first part is clear. As for the second, we have to go from the
right action to the left action: In general
if $x\in\g(U_S^\times)$, let us define the right derivative
\be
D_x^ru(\hat g)=\dds u(\hat g\, [(\exp(sx_{\mathrm{reg}}),1)]),
\ee
for any smooth $x_{\mathrm{reg}}$ on $U_S$ equal to $x$ away from
a neighborhood of $S$. As for the left derivative, this object is
independent of the choice of regularization.

\begin{lemma}\label{llr}
For any $u\in E_{V,k}(\Sigma,S)$, $x\in\g(U_S^\times)$ and
$\hat g\in\hat G(U_S)^\times)$ with $\pi(\hat g)=g$,
\be
D^r_xu(\hat g)=-D_{\Ad(g_\lambda)x}u(\hat g)-
\frac k{2\pi i}\oint_\gamma\tr(xg^{-1}dg)
\ee
\end{lemma}
\begin{prf}
Let $\hat g=[( g_{\mathrm{reg}},w)]$, for some smooth $ g_{\mathrm{reg}}$
 on $U_S$.
Let $h_{\mathrm{reg}}=\exp(s x_{\mathrm{reg}})$, where $x_{\mathrm{reg}}$
is a smooth map on $U_S$ which coincides with $x$ on the
complement of a small neighborhood of $S$. 
\begin{equation}\label{aaa}
u([(g_{\mathrm{reg}},w)][(h_{\mathrm{reg}},1)])
=
u([(g_{\mathrm{reg}} h_{\mathrm{reg}} 
g_{\mathrm{reg}}^{-1},1)]\,\hat g)
e^{k\Gamma(g_{\mathrm{reg}},h_{\mathrm{reg}})},
\end{equation}
with
\bea
\Gamma(g,h_{\mathrm{reg}})&=&
\frac 1{2\pi i}\int_{U_S}
(\omega(g_{\mathrm{reg}},h_{\mathrm{reg}})-
\omega(g_{\mathrm{reg}} h_{\mathrm{reg}}
 g_{\mathrm{reg}}^{-1},g_{\mathrm{reg}}))
\\
&=&-s\frac 1{2\pi i}\oint_\gamma\tr(g^{-1}dg\,x)+O(s^2).
\eea
Now, $g_{\mathrm{reg}} x_{\mathrm{reg}}
g_{\mathrm{reg}}^{-1}$ is smooth on $U_S$
and coincides with $\Ad(g)x$ on the complement in $U_S$ of a neighborhood
of $S$. Taking the derivative at $s=0$ of \Ref{aaa}, we obtain the
result.
 \end{prf}

In particular this completes the proof of
the second part of the preceding Lemma:
take $x=\Ad(g^{-1})y$, and use the fact that, by right invariance,
$D^r_xu=0$, if $x\in\g(\Sigma-S)$. The $WZW$ action in Theorem
\ref{subgroups}, (ii) does not enter here, since it vanishes to
second order in $s$ if $h=\exp(sx)$.

The next important fact about conformal blocks is that if the representation
associated to a point $p\in S$ is trivial, then the space of conformal
blocks is canonically identified with the space associated to $S-\{p\}$.
Consequently, we may think of conformal blocks as taking values in
$\otimes_{x\in \Sigma}V_x$ where all but finitely many $V_x$ are trivial.

\begin{lemma}\label{l5}
Let $R\subset S$ be a non-empty subset and
suppose $V_j=\C$, the trivial representation, if $p_j\not\in R$.
 Let us embed
$G(U^\times_R)$ into $G(U_S^\times)$ by extension
by $1$.  This embedding lifts uniquely to an embedding
$i:\hat G(U^\times_R)\hookrightarrow \hat G(U_S^\times)$,
preserving the central subgroup $\C^\times$.
Let $V'=\otimes_{p_j\in R}V_i$. Then the pull-back 
$i^*:u\mapsto u\circ i$ is an isomorphism from
$E_{V,k}(\Sigma,S)$ onto $E_{V',k}(\Sigma,R)$.
\end{lemma}

\section{The connection}

\subsection{The energy-momentum tensor}

The construction of the connection involves the introduction
of differential operators associated to an additional point $p$
which varies on the complement of $S$.

Let $U_0$ be some embedded disk  in $\Sigma$, disjoint from $U_S$,
and choose a local 
coordinate $z:U_0\to\C$. Let $p\in U_0$, and set $U_{S\cup\{p\}}=
U_S\cup U_0$. For any $x\in\g$ let 
$x_j(p)\in\g(U_{S\cup\{p\}}^\times)$ be the map $q\mapsto x(z(q)-z(p))^j$,
if $q\in U_0$ and $q\mapsto 0$ if $q\in U_S$.
Denote by $i^*_p$ the isomorphism
$E_{V,k}(\Sigma,S\cup\{p\})\to E_{V,k}(\Sigma,S)$ of Lemma \ref{l5}.
We define a differential operator
$J_x(p)$, depending linearly on $x\in\g$ and acting on conformal 
blocks:
\be
J_x(p)u(\hat g)=
D_{x_{-1}(p)}i_p^{*-1}u(i_p(\hat g)), \qquad 
\hat g\in \hat G(U_S^\times).
\ee
This differential operator depends on the choice of local coordinate
$z$. However the 1-form $J_x(p)dz(p)$ does not when acting on
conformal blocks:
\begin{lemma}
Let $J^z_x(p)$, $J^w_x(p)$ be the above differential operators defined
using the local coordinates $z$ and $w$ respectively. Then for
all $p$ in the common domain of definition and 
$u\in E_{V,k}(\Sigma,S)$,
$J_x^z(p)u(g)=(dw/dz)(z(p))J^w_x(p)u(g)$.
\end{lemma}
\begin{prf}
We have
\be
\frac x{z(q)-z(p)}=\frac x{w(q)-w(p)}\;\frac{dw}{dz}(z(p))+\cdots,
\ee
where the dots stand for a function of $q$ which is regular at $p$.
This regular function does not contribute when acting on $E_{V,k}(\Sigma,S)$.
Thus if $u\in E_{V,k}(\Sigma,S)$,
then $J^z_x(p)u=J^w_{\tilde x}(p)u$ with $\tilde x=x(dw/dz)(z(p))$.
The claim follows by linearity.
\end{prf}

Let us fix some local coordinate on $U_0$ and write simply $J_x(z)dz$
for $J_x(p)dz(p)$. Also, let $J(z)dz$ be the linear function
on $\g$ sending $x$ to $J_x(z)dz$.

\begin{proposition}\label{wherestheproof}
Let $u\in E_{V,k}(\Sigma,S)$ and $\hat g\in \hat G(U_S^\times)$ with
$\pi(\hat g)=g$. Then
\begin{enumerate}
\item[(i)]
The one-form $J(z)u(\hat g)dz$ on $U_0$, with values in $\g^*\otimes V$,
has an analytic continuation to a one-form still denoted $J(z)u(\hat g)dz$ 
on $\Sigma-S$, and $J(z)$ is a first order differential operator
on conformal blocks.
Moreover, for all $x\in\g$, the differential operator
\be
J_x^S(z)u(\hat g)dz=J_{\Ad(g(z)^{-1})x}(z)u(\hat g)dz
-k\,\tr (x dg(z)g(z)^{-1})u(\hat g),
\ee
defined on $U_S^\times$, extends to a meromorphic function of
$z$ with at most simple poles on $S$.
\item[(ii)]
If $y\in\g(U_S^\times)$, then
\bea
D_y^ru(\hat g)&=&-\frac 1{2\pi i}
\oint_\gamma\langle y(z),J(z)\rangle u(\hat g)dz,
\\
D_yu(\hat g)
&=&\frac 1{2\pi i}
\oint_\gamma\langle y(z),J^S(z)\rangle u(\hat g)dz.
\eea
\end{enumerate}
\end{proposition}

The proof is deferred to \ref{herestheproof}, after we have introduced
a more explicit description of how $J(z)$ acts on conformal blocks.
At this point we only remark that by Lemma \ref{llr}, the two
formulae in (ii) are equivalent.

Now let $b_1,\dots, b_D$ be a basis of $\g$ such that $\tr(b_ib_j)=
\delta_{ij}$. Let $h^\vee=N$ 
be the dual Coxeter number of $G$.
The {\emph{energy-momentum tensor}} is the differential
operator
\be
T(p)u(g)=\frac1{2(k+h^\vee)}\sum_{j=1}^DD_{(b_j)_{-1}(p)}^2
i_p^{*-1}u(i_p(g)).
\ee
Naively, $T(p)$ should transform as a quadratic differential under
change of coordinate. However there is a correction term, due to
the fact that after the first application of $D_{b_j}$ the resulting
function is not in general
a conformal block. We proceed to give a formula for this correction
term.
 If $w(z)$ is a holomorphic function
of a complex variable, the {\emph{Schwarzian derivative}} of $w$ is
the function 
\be
\{w,z\}=\frac {w'''(z)}{w'(z)}-\frac32\,\left(\frac{w''(z)}{w'(z)}
\right)^2.
\ee
Its main properties are the chain rule
\begin{equation}\label{Sch}
\{w,z\}\left(\frac{dz}{du}\right)^2+\{z,u\}=\{w\circ z,u\}
\end{equation}
and the fact that $\{w,z\}=0$ if and only if $w(z)$
has the form $\frac{az+b}{cz+d}$, see,  e.g., \cite{T}.
\begin{lemma}\label{lS}
If $u\in E_{V,k}(\Sigma,S)$ and $T^z(p)$, $T^w(p)$ are the differential
operators $T(p)$ defined using local coordinates $z$, $w$, 
respectively, then
\be
T^z(p)\,u=\left(\frac{dw}{dz}(z(p))
\right)^2T^w(p)\,u+\frac1{12}c_k\{w,z(p)\}\,u,\qquad
 c_k=\frac {k D}{k+h^\vee}
\ee
where $D=N^2-1$ is the complex dimension of $G$ and $h^\vee=N$
is the dual Coxeter number of $\g$.
\end{lemma}
\begin{prf}
Let $w(z)$ be the function expressing the coordinate $w$ in terms
of the coordinate $z$, and let us write $z$ instead of $z(q)$ and
$z_0=z(p)$. Then we have the expansion in powers of
$t=z-z_0$.
\be
\frac{w'(z_0)}{w(z)-w(z_0)}=
\frac1{t}
-
\frac{w''(z_0)}{2w'(z_0)}
-\frac16\{w,z_0\}t
+O(t^2).
\ee
Correspondingly, for any $x\in\g$, we have, indicating the
choices of coordinate as superscripts,
\be
w'(z_0)x^w_{-1}(p)=x^z_{-1}(p)-
\frac{w''(z_0)}{2w'(z_0)}x^z_0(p)
-\frac16\{w,z_0\}x^z_1(p)
+r_x
\ee
and $r_x$ vanishes to second order at $p$.
Now if $u$ is a conformal block, then $D_{x_j(p)}u=0$ for all $j\geq 0$,
by invariance under $G(U_0)\subset \hat G(U_{S\cup\{p\}}^\times)$. Thus
we have (the derivatives are taken at $z_0$)
\bea
{w'}^2D_{x^w_{-1}(p)}^2u(g)&=&
(D_{x^z_{-1}(p)}-\frac{w''}{2w'}D_{x^z_0(p)}
-\frac16\{w,z\}D_{x^z_{1}(p)})D_{x^z_{-1}(p)}u(g)\\
&=&
(D_{x^z_{-1}(p)}^2-\frac{w''}{2w'}[D_{x^z_0(p)},D_{x^z_{-1}(p)}]\\
 & &-\frac16\{w,z\}[D_{x^z_{1}(p)},D_{x^z_{-1}(p)}])u(g)\\
&=&
D_{x^z_{-1}(p)}^2u(g)-\frac16\{w,z\}\,k\,\tr(x^2)u(g).
\eea
By taking $x=b_j$,  summing over $j$ and dividing by
$2(k+h^\vee)$ we get the desired result.
\end{prf}

We will need to know the behavior of $T(p)u$ under the action of
$G(U_S)$ and $G(\Sigma-S)$.
\begin{lemma}\label{lemma34} For all $\hat g\in\hat G(U_S^\times)$,
\begin{enumerate}
\item[(i)]
$T(p)u(b\hat g)=b\,T(p)u(\hat g)$, if $b\in G(U_S)$.
\item[(ii)]
$
\frac d{ds}\big|_{s=0} T(p)
u(\hat g\exp(sx))=J_{x'(p)}u(\hat g)$, if $x\in\g(\Sigma-S)$
\end{enumerate}
Here $x'(p)$ denotes the derivative of $x$ at $p$ with respect to the
local coordinate used to define $T$.
\end{lemma}
\begin{prf}
(i) is obvious. To prove (ii), we use the invariance of $i_p^{*-1}u$ under
the right action of $G(\Sigma-(S\cup\{p\}))$: let $y(t)=x(t)$ for 
$t\in U_0$ and $y(t)=0$ for $t\in U_S$. Then $i_p(\hat g \exp(sx))=
i_p(\hat g)\exp(-sy)\exp(sx+O(s^2))$. Therefore, if $\kappa=k+h^\vee$,
\bea
2\kappa\dds T(z)u(\hat g\exp(sx))
\!\!\!&=&\!\!\!
\dds\sum_{j=1}^D
D_{(b_j)_{-1}(p)}^2i_p^{*-1}u(i_p(\hat g\exp(sx)))
\\
&=&\!\!\!
\sum_{j=1}^D
D_yD_{(b_j)_{-1}(p)}^2i_p^{*-1}u(i_p(\hat g)).
\eea
The last operator can be replaced by the commutator
$[D_y,D_{(b_j)_{-1}(p)}^2]$, since $D_yi_p^{*-1}u=0$ by the
regularity of $y$ on $U_0$. This commutator can be computed
using Proposition \ref{propLie}. The four properties one has to
use are: the invariance of $C=\sum b_j\otimes b_j$, i.e.,
$[C,x^{(1)}+x^{(2)}]$ for any $x\in\g$;  the fact that
$h^\vee$ is half the Casimir of the adjoint representation,
i.e., $\sum_j\ad(b_j)\ad(b_j)=2h^\vee\Id_\g$; the invariance
of $\tr$, i.e., $\tr([x,y]z)=\tr(x[y,z])$, in particular
$\tr([x,b_j]b_j)=0$; $D_yi_p^*u=0$ if $y$ is zero on $U_S$ and
regular on $U_0$. The result is
\be
\sum[D_y,D^2_{(b_i)_{-1}(p)}]i_p^{*-1}u=
D_ai_p^{*-1}u,
\ee
where $a(q)=x'(p)/(z(q)-z(p))$.
\end{prf}

\subsection{Flat structures}
We follow \cite{T}. A {\emph{flat structure}} on a compact Riemann surface
$\Sigma$ is an equivalence class of flat atlases 
$\{(U_\alpha,z_\alpha), \alpha\in I\}$. A flat atlas is 
covering of $\Sigma$ by open sets $U_\alpha$ with local coordinates
$z_\alpha:U_\alpha\to\C$ such that the transition functions
$z_\alpha\circ z_\beta^{-1}$ are M\"obius transformations. Two
flat atlases are equivalent if their union is a flat atlas.

Flat structures exist on any $\Sigma$ 
by the Riemann uniformization theorem. The description  of
all the flat structure on a  compact Riemann surface $\Sigma$  is
given by 
the following result.
\begin{thm} (See \cite{T})
The set of flat structures on a compact Riemann surface
$\Sigma$ is an affine space over the
vector space $H^0(\Sigma,K^2)$ of quadratic differentials.
\end{thm}
If we have two flat structures given  by local coordinates
$z_\alpha$, $w_\alpha$, respectively (we may assume that they
are defined on the same covering by going to a refinement),
the associated quadratic differential (their difference in the 
affine space)
is given on $U_\alpha$
by the Schwarzian $\{z_\alpha,w_\alpha\}dw_\alpha^2$. The
properties \Ref{Sch} of the Schwarzian derivative ensure that
the quadratic differential is defined globally, independently
of the choice of atlas within its equivalence class, and that
this defines an affine structure on the set of flat structures.

Thus if we fix a flat structure and consider only coordinates
in a flat atlas, we see that the energy momentum tensor depends
on coordinates like a quadratic differential.

\subsection{Connections on bundles of projective spaces}\label{cobops}
The Friedan--Schenker connection is a connection on the bundle of
projective spaces of conformal blocks over the moduli space of curves
(with additional data).

This means the following: if $E$ is a holomorphic vector bundle
over a complex manifold $X$, let $\Pee(E)$ be the bundle
of projective spaces whose fiber at $x$ is the projectivization
$\Pee(E_x)$ of the fiber $E_x$. A connection on $\Pee(E)$ is
an equivalence class of locally defined connections $\nabla_\alpha$
on the holomorphic vector bundles $E|_{U_\alpha}$,
 for some open covering $(U_\alpha)$, such that on intersections
$U_{\alpha,\beta}=
U_\alpha\cap U_\beta$, $\nabla_\alpha-\nabla_\beta$ is a {\em scalar}
holomorphic 1-form $a_{\alpha,\beta}\in H^0(U_{\alpha\beta}, T^*M)$.
 Two connections
are equivalent if their difference is locally a scalar holomorphic one-form. 
Such data define a connection on $\Pee(E)$ globally: for each
curve $t\mapsto\gamma(t)$ in $M$, the lift $\tilde\gamma(t)={\mathrm{cls}}(
s(\gamma(t)))$ with $\nabla_{\dot\gamma}s(\gamma(t))=0$, is uniquely
and unambiguously determined by the initial condition $\tilde\gamma(0)$. 

The curvature
of a connection on $\Pee(E)$ is locally a two-form $F_\alpha=\nabla_\alpha^2$
with values in $\End(E)$ such that, on $U_{\alpha\beta}$,
$F_\alpha-F_\beta=d a_{\alpha\beta}$. Equivalent connections give rise
to curvatures that differ locally by exact scalar one-forms.

A connection on $\Pee(E)$ is flat if $F_\alpha$ is a {\em scalar} (i.e., taking
its values in the trivial subbundle of $\End(E)$ consisting of multiples
of the identity) one-form. This means
that contractible closed curve are lifted to closed curves in $\Pee(E)$.

There are two points of view when considering flat connections on
$\Pee(E)$: the \v Cech and the Dolbeault point of view, in Hitchin's
terminology \cite{H1}. The curvature of a flat connection is given
by a set of closed two-forms $F_\alpha$ defined up to addition of
exact forms, and such that $F_\alpha-F_\beta$ is exact on intersections.
In the \v Cech description, we use the fact that locally $F_{\alpha}$ is
exact to  choose a representative $\nabla_\alpha$ with $F_\alpha=0$,
so that a flat connection on $\Pee(E)$ is described by genuinely flat,
locally defined connections on $E$. In the Dolbeault description,
we may find smooth one-forms $c_\alpha$ such that $a_{\alpha\beta}=
c_\alpha-c_\beta$. Then $\nabla_\alpha+c_\alpha$ is a globally
defined connection on the vector bundle, but has non-trivial 
(scalar) curvature.

In our situation, the connection is described in the following terms.
Let us fix the two-dimensional oriented
smooth compact manifold $\Sigma$ of genus $h$ with marked points 
$S=\{p_1,\dots, p_n\}$. Fix also non-zero tangent vectors $v_1,\dots
v_n$ at the points. Consider the moduli space 
$\hat M_{h,n}$ of complex structures $I$ modulo 
the action of the group $\Diff$ of orientation preserving diffeomeorphisms
which leave the points and the tangent vectors fixed.
This moduli space has a
natural structure of algebraic variety of dimension
$3({\mathrm{genus}}(\Sigma)-1)+2n$, provided the genus is at least two,
or one and $n\geq 1$, or zero and $n\geq2$.

A tangent vector to $ \hat M_{h,n}$ is a class in
$H^{1}(\Sigma,T\Sigma\otimes O(-2S))$ by the Kodaira--Spencer
deformation theory ($S$ is identified with the divisor $p_1+\cdots+p_n$).
If we choose local coordinates on a neighborhood of $S$, a
class $[\zeta]$ is represented, in the \v Cech formulation,
 by a holomorphic vector
field $\zeta(z)\ddz$ on $U_S^\times$, modulo vector fields
extending to $\Sigma-S$ and vector fields extending to $U_S$
and vanishing at $S$ to first order%
\footnote{We say that a
that a  vector field  vanishes to first order at
a point $p$, if its components vanish and have vanishing first derivative,
for some, and thus any, choice of coordinates at $p$. A vector
field vanishes to first order at a set $S$ if it vanishes to 
first order at all its points.}.
If we view $I(p)$ as an endomorphism of the cotangent space 
at $p$ with
$I(p)^2=-1$, then an infinitesimal deformation $\dot I$ of $I$ is 
described by a Beltrami differential 
$\mu=\mu(w)\frac\partial{\partial w}\otimes d\bar w$, whose
local coordinate expression is defined by $\dot I\,dw=2i\mu(w)d\bar w$.
The Beltrami differentials define the complex structure on the
infinite dimensional manifold of complex structures on $\Sigma$: 
$\mu$ is the $(1,0)$ component of the tangent vector $\dot I$.
The connection with the \v Cech  description is that
on $U_S$, $\mu=\bar\partial \zeta_S$ and on $\Sigma-S$, $\mu=\bar
\partial\zeta_\infty$
for some smooth vector fields $\zeta_S$, $\zeta_\infty$, defined
up to addition of holomorphic vector fields such that
$\zeta_S$ vanishes to first order at $S$. The difference
$\zeta=\zeta_S-\zeta_\infty$ is holomorphic on $U_S^\times$.

The Friedan--Shenker connection is a connection on the
projectivized vector bundle  of conformal blocks over
$\hat M_{h,n}$. Its fiber over $[I]$ is the projectivized vector space
of conformal blocks $\Pee E_{V,k}(\Sigma,I,S)$, where we indicate
the dependence on the complex strucure in the notation,
for any choice of representative $I$. Conformal blocks associated
to equivalent complex structures are canonically identified.
A local  holomorphic section of this bundle is represented by a holomorphic
function $I\mapsto u(I,\hat g)$ invariant under the natural action
of $\Diff$, and such that $u(I,b\hat g n)=bu(I,\hat g)$
and $u(I,gz)=z^ku(I,g)$, see \ref{scb}, for all $b(I,t)$ 
 Given a local holomorphic section $u$,
and a local holomorphic vector field $\zeta$, we define
the covariant derivative $\nabla_\zeta u$. To define
the value of $\nabla_{\zeta}u$ at $\hat g$ in concrete terms,
 we need to specify how $\hat g$ changes as we deform the complex structure
to take the derivative.
For this we choose local coordinates to identify $U_S$ with
a fixed union of disks, so that an element $\hat g\in \hat G(U_S^\times)$
is defined even if we vary the complex structure.
Thus we consider an enlarged moduli space of data $(I,z)$,
where $I$ is again a complex structure and $z$ is a local coordinate
$z:U_S\to\sqcup_{j=1}^n\C$ defined on some neighborhood $U_S$,
mapping $p_j$ to the origin of the
$j^{\mathrm{th}}$ copy of $\C$, and such that the tangent map
$z_*(p_j)$ sends $v_j$ to $1$.
The equivalence relation is given as before by the
action of $\Diff$.
 We have the natural projection $(I,z)\mapsto I$ from the enlarged
moduli space to $\hat M_{h,n}$,
with contractible fiber.
 
 Let us now describe the tangent space at a point of the
enlarged moduli space.
If $z$ is a complex coordinate for $I$ on $U_S$ we have by definition
$I\,dz=i\,dz$. Differentiating, we see that a
 tangent vector $(\dot I,\dot z)$
at $(I,z)$ is  then given by a pair obeying
$\dot I dz=i(1+iI)d\dot z$
 or $\mu(z)=\frac\partial{\partial\bar z}\dot z$ on $U_S$.
Thus a tangent vector $(\dot I,\dot z)$ gives rise to
a vector field $\zeta=\zeta_S-\zeta_\infty$ on $\g(U_S^\times)$,
with $\zeta_S=\dot z$, defined up to addition of
a holomorphic
vector field on $\Sigma-S$. Conversely, given a holomorphic
vector field $\zeta$ on $U^\times_S$, we may
write $\zeta=\zeta_S-\zeta_\infty$ as the difference of 
two smooth vector fields extending to $U_S$, $\Sigma-S$,
respectively, and such that $\zeta_S$ vanishes to first order
at $S$.
 This defines a tangent vector $(\dot I,\dot z)$
with $\mu=\bar\partial \zeta_S$ on $U_S$, $\mu=\bar\partial\zeta_\infty$
on $\Sigma-S$ and $\dot z=\zeta_S$. Different choice of
$\zeta_S$, $\zeta_\infty$ lead to tangent vectors differing by
infinitesimal diffeomorphisms.

In other words, a tangent vector to the enlarged moduli
space is the same as a holomorphic
vector field $\zeta(z)\ddz$ on $U_S^\times$
defined modulo  holomorphic  vector fields extending to $\Sigma-S$.

Let $u(I,\hat g)$ be a local holomorphic section of the vector bundle
of conformal blocks. Thus $u$ is a holomorphic map $I\mapsto
u(I,\cdot)\in E_{V,k}(\Sigma,I,S)$, invariant under the natural
action of $\Diff$.

To define the covariant derivative in the direction of a tangent
vector $[\zeta]$, we choose a one parameter family
 $(I_\tau,z_\tau)$, $|\tau|<\epsilon$,
be a holomorphic  of data in the enlarged moduli space, so that
the class of $\dot I=\frac d{d\tau}|_{\tau=0}I_\tau$ is $[\zeta]$. 
If $\hat g=[(g_{\mathrm{reg}},1)]$ is in $\hat G$ for 
the complex structure $I=I_0$, then $\hat g_\tau=
[(g_{\mathrm{reg}}\circ z^{-1}\circ z_\tau,1)]$ is in $\hat G$ for
the complex structure $I_\tau$ (its expression in the local coordinate
$z_\tau$ coincides with the expression of $\hat g$ in terms of the
local coordinate $z$).
 Let $\zeta$ be
a vector field corresponding to $(\dot I_\tau,\dot z_\tau)$
at $\tau=0$.

The covariant derivative is defined by  an expression
\begin{equation}\label{e300}
\nabla_{\zeta} u(I,\hat g)=
\frac d{d\tau}\bigg|_{\tau=0}u(I_\tau,\hat g_\tau)
+A(I,\zeta)u(I,\hat g),\qquad z=z_{\tau=0}.
\end{equation}
Here $A(I,\zeta)$ is a differential operator on conformal blocks
depending linearly on $\zeta$. In order for this formula
to define a connection on $\hat M_{h,n}$ we need to show
that 
\begin{equation}\label{e323}
 A(I,\eta)=0,
\end{equation}
 if $\eta$ extends to a holomorphic vector field
on $\Sigma-S$ and that if $\eta$ is holomorphic on $U_S$
and vanishes to first order at $S$, then
\begin{equation}\label{e324}
A(I,\eta)u(\hat g)=D_{\eta g'g^{-1}}u(\hat g)
+\frac k{4\pi i}\oint_\gamma\eta(z)\tr((g'(z)g(z)^{-1})^2)dz 
\,u(\hat g)
\end{equation}
Here $g=\pi(\hat g)$. 

The origin of \Ref{e324} is the following: if we replace 
$z_\tau$ by some other coordinate $w_\tau$ such that
$w_0=z_0$ and such that $\dot w_0-\dot z_0=\eta$, then
the tangent vector $\zeta$ is replaced by $\zeta+\eta$.
the right-hand side of \Ref{e300} is replaced by
\begin{equation}
\frac d{d\tau}\bigg|_{\tau=0}u(I_\tau,[(g_{\mathrm{reg}}\circ z^{-1}
\circ w_\tau)])
+A(I,\zeta+\eta)u(\hat g),
\end{equation}
which is the same as \Ref{e300}, if we have
\Ref{e324}, as a straightforward calculation shows.

Summarizing, we have:

 \begin{proposition}\label{pA}
Let for $I$ in some open set of the space of complex structures
on $\Sigma$, and $\zeta$ a holomorphic vector field
defined on some pointed neighborhood of $S$, $A(I,\zeta)$
 be a differential operator acting
on $E_{V,k}(\Sigma, I,S)$,
depending linearly on $\zeta$. Then $A(I,\zeta)$
defines locally via \Ref{e300} a connection on the bundle of projective
spaces of conformal blocks, if (i) $\nabla_\zeta$ maps local
sections to local sections, (ii) $A(I,\zeta)=0$ if $\zeta$ extends
to a holomorphic vector field on $\Sigma-S$ and 
(iii) eq.~\Ref{e324} holds if $\eta$ is regular at $S$ and vanishes
there to first order.
\end{proposition}

\subsection{The Friedan--Shenker connection}
We now claim that the energy momentum tensor can be used
to construct a differential operator $A$ obeying the hypotheses
of Proposition \ref{pA}, and thus defines a connection on 
$M_{h,n}$.
\begin{proposition}\label{wheresitsproof} Let $\kappa=k+h^\vee$.
  Let us fix a flat structure on $(\Sigma, I)$.  and let $z$ be a
  coordinate on $U_0$. Then $T(p)dz(p)^2$, defined for $p\in U_0$,
  extends, as a function of $p$, to a differential-operator-valued
  quadratic differential still denoted $ T(p)dz(p)^2$
  on $\Sigma-S$. Moreover, if $p\in U_S^\times$, 
  \bea
   T^S(p)dz(p)^2u(\hat g)\!\!\!&=&\!\!\!\big(T(p)dz(p)^2+
   \tr^{(0)}\big((g(p)^{-1}dg(p))^{(0)}
   J(p)\big)dz(p)
   \\
   & & +\frac
   k2\tr((g^{-1}(p)dg(p))^2)\big)u(\hat g),
  \eea
in local coordinates on $U_S$, extends to a meromorphic quadratic
differential on $U_S$ with at most poles of second order on $S$.
Here the square in the last term is the symmetric square, sending
differentials to quadratic differentials.
\end{proposition}
The proof is deferred to \ref{herestheproof}.

\begin{thm}\label{fsc} Let $T(z)dz^2$ be a local coordinate
expression, for some choice of flat structure,
 of the energy momentum tensor and $\zeta(z)\ddz$ a holomorphic
vector field on $U_S^\times$.
Then
$A(I,\zeta)=\frac1{2\pi i}\oint_\gamma\zeta(z)T(z)dz$
defines locally, via \Ref{e300}, a connection on the vector 
bundle of conformal blocks.
 \end{thm}

\begin{prf}We show that the criteria of Proposition \ref{pA}
are satisfied.
We first show,
using  Lemma \ref{lemma34} and Proposition \ref{wherestheproof} (ii),
that
the connection is well-defined, i.e., the
covariant derivative maps local sections to local sections.
Clearly $\nabla_\zeta u(b\hat g)=b\nabla_\zeta u(\hat g)$ 
by Lemma \ref{lemma34} (i). Let $n\in G(\Sigma-S)=G(\Sigma-S,I)$, for the
complex structure $I$, and suppose $n$ is part of a holomorphic
family $\tilde n_\tau$, such that $\tilde n_\tau\in G(\Sigma-S,I_\tau)$. 
We must compare this deformation with the deformation $n_\tau=n\circ
z^{-1}\circ z_\tau$ appearing in the definition of the connection.
We
have, by the invariance of $u$,
\bea
\frac d{d\tau}\bigg|_{\tau=0}u(I_\tau,(\hat g n)_\tau)
&=&
\frac d{d\tau}\bigg|_{\tau=0}u(I_\tau,\hat g_\tau n_\tau)
\\
&=&
\frac d{d\tau}\bigg|_{\tau=0}u(I_\tau,\hat g_\tau n_\tau\tilde n_\tau^{-1})
\\
&=&
\frac d{d\tau}\bigg|_{\tau=0}u(I_\tau,\hat g_\tau)
+D^r_{\zeta n'n^{-1}}u(I,\hat g).
\eea
In the last term the prime means the derivative with respect to 
the local coordinate $z$ on $U_S$. The point is that the fact that
$\tilde n_\tau$ extends to $\Sigma-S$ implies that 
\be
\frac\partial{\partial\tau}\bigg|_{\tau=0}\tilde n_\tau+\zeta n'
\ee
extends to a holomorphic function on $\Sigma-S$ with values in $\g$
for the complex structure
$I$. This translates into the necessary condition for $A$ to
define a connection:
\be
A(I,\zeta)u(\hat g n)=A(I,\zeta)u(\hat g)-D^r_{\zeta n'n^{-1}}u(I,\hat g),
\ee
or, infinitesimally ($G(\Sigma-S)$ is connected),
\be
\dds A(I,\zeta)u(\hat g\exp(sx))=-D^r_{\zeta x'}u(I,\hat g),
\ee
for all $x\in\g(\Sigma-S)$.
This property follows from Lemma \ref{lemma34} (ii) with
Proposition \ref{wherestheproof}.

The property (ii) of Proposition \ref{pA} follows from 
the fact stated in Proposition \ref{wheresitsproof} that
$T(z)dz^2$ is a holomorphic quadratic differential on 
$\Sigma-S$, by Stokes' theorem.

By the same argument, we have $\oint T^S(z)\eta(z)dz=0$, if
$\eta(z)d/dz$ is holomorphic on $U_S$. Thus, by the formula
in Proposition \ref{wheresitsproof}, we obtain
\bea
0&=&A(\eta)+\frac1{2\pi i}\oint_\gamma\langle(g^{-1}dg,J\rangle\eta
+\frac k{4\pi i}\oint_\gamma\eta(z)\tr((g'g^{-1})^2)dz 
\\  &=&A(\eta)+\frac1{2\pi i}\oint_\gamma\langle dg g^{-1},J^S\rangle\eta
-\frac k{4\pi i}\oint_\gamma\eta(z)\tr((g'g^{-1})^2)dz 
\\ &=&A(\eta)-D_\eta g'g^{-1}
-\frac k{4\pi i}\oint_\gamma\eta(z)\tr((g'g^{-1})^2)dz,
\eea
which is (iii) of Proposition \ref{pA}.
\end{prf}

\noi{\bf Remark.}\ The dependence on the choice of flat structure
is given by Lemma \ref{lS}. Thus if we change flat structure
we get an equivalent connection in the sense of \ref{cobops}.

\section{The Knizhnik--Zamolodchikov--Bernard equations}

We give here formulae for the connection. We assume that
the genus of $\Sigma$ is at least two. The cases of genus
zero and one require slight modifications.

\subsection{Dynamical $r$-matrices}\label{drm}
In this section, some of the results of \cite{EF} are reviewed.
The setting is the same as in Section \ref{sec3}. Furthermore, we
assume that the genus of $\Sigma$ is at least two.
\begin{proposition}\label{prop12}
Let $g\in G(U^\times)$ be such that
$H^0(\Sigma,{\mathrm{Ad}}(P_g))=0$. Let $C$ be the invariant
symmetric tensor $\sum_jb_j\otimes b_j$ for any basis
basis $(b_j)$ of $\g$ so that $\tr(b_ib_j)=\delta_{ij}$.
Let $\xi_1,\dots,\xi_m\in\g(U_S^\times)$ be representatives of
a basis of \v Cech cohomology classes in 
$H^1(\Sigma,{\mathrm{Ad}}(P_g))$, $m=({\mathrm{genus}}(\Sigma)-1)\dim(G)$.
Then, for each fixed $w\in U_S$ there exists a unique meromorphic
one-form $r(z,w)dz$ on $U_S$ with values in $\g\otimes\g$ such that
\begin{enumerate}
\item[(i)] $r(z,w)dz$ is regular on $U_S-\{w\}$ and, as $z\to w$,
\be
r(z,w)dz=\frac C{z-w}\,dz + O(1),
\ee
\item[(ii)] 
$\Ad(g(z)^{-1})^{(1)}r(z,w)dz$ extends to a holomorphic one-form
on $\Sigma-S$.
\item[(iii)] Let $\gamma$ be the sum of simple closed curves 
in $U_S$ encircling counterclockwise 
the points $p_1\dots,p_n\in S$ and $w$.
Then 
\be
\oint_\gamma\tr^{(1)}\left( \xi_j(z)^{(1)} r(z,w)\right)dz=0,
 \qquad j=1,\dots,m
\ee
\end{enumerate}
\end{proposition}

We use here the notation $x^{(j)}$ to denote the action of the
linear map $x$ on the $j^{\mathrm {th}}$ factor of a tensor product
of vector spaces. So, for instance, $\tr^{(1)}(x\otimes y)=\tr(x)\,y$.

This object $r(z,w)dz$, the classical $r$-matrix, appears in \cite{EF}
in a description of Poisson brackets and integrals of motions of
Hitchin systems. 

\begin{lemma}\label{transf}
Let $r(z,w;g,(\xi_j))$ be the classical  $r$-matrix corresponding
to the
data $g\in G(U_S^\times)$ and $\xi_1,\dots, \xi_m\in\g(U_S^\times)$. 
Then, for any $h\in G(U_S)$,
\be
r(z,w;hg,(\Ad(h)\xi_j))=\Ad(h(z))\otimes\Ad(h(w))
r(z,w;g,(\xi_j)).
\ee
For any $n\in G(\Sigma-S)$, $r(z,w;gn,(\xi_j))=r(z,w;g,(\xi_j))$.
For any $x_1,\dots,x_m\in \g(U_S)$, 
\be
r(z,w;g,(\xi_j+x_j))=r(z,w;g,(\xi_j))-\sum_{j=1}^m\omega_j(z)\otimes x_j(w).
\ee
Finally, for any $y_1,\dots,y_m\in\g(\Sigma-S)$,
\be
r(z,w;g,(\xi_j+\Ad(g)y_j))=r(z,w;g,(\xi_j)).
\ee
\end{lemma}
\begin{prf}
By the uniqueness of $r$,
the proof consists of checking the characterizing properties
(i)-(iii), which is straightforward.
\end{prf}
The dependence of $r(z,w)$ on the second argument is given by
the following result. Let
$\omega_j(z)dz$, $j=1,\dots,m$,
 be a basis of one-forms in $H^0(\Sigma,K\otimes\Ad(P_g))$
dual (by Serre duality) to the basis $(\xi_j)$. 

Thus $\omega_j(z)dz$ is
a $\g$-valued holomorphic one-form on $U_S$,
 such that $\Ad(g(z)^{-1})\omega_j(z)dz$
extends to a holomorphic one-form on $\Sigma-S$ and
\be
\frac 1{2\pi i}\oint\tr\,\omega_j(z)\xi_l(z)dz=\delta_{jl}.
\ee
\begin{lemma}\label{l99}
For fixed $z\in U_S$, the classical $r$-matrix $r(z,w)$ is a holomorphic
function of $w\in U_S-\{z\}$. Moreover $\Ad(g(w)^{-1})^{(2)}
(r(z,w)+\omega_j(z)\otimes\xi_j(w))$
 extends to a holomorphic function of $w$ on $\Sigma-(S\cup\{z\})$
\end{lemma} 

A different characterization of classical $r$-matrices is as
kernels of projections (cf.~\cite{C}, \S 1). In the present context,
this characterization also holds:
\begin{thm}\label{proj}
Let $\g(\Sigma-S,g)$ be the Lie subalgebra of $\g(U_S^\times)$ of
functions $x$, such that $\Ad(g^{-1})x$ extends to $\Sigma-S$.
Let $P_+:\g(U_S^\times)\to\g(U_S^\times)$ be the projection
onto $\g(U_S)$ in the decomposition
\begin{equation}\label{decomposition}
\g(U_S^\times)=\g(U_S)\oplus(\oplus_{j=1}^{m}\C\,\xi_j)\oplus 
\g(\Sigma-S,g).
\end{equation}
Then, for any $x\in\g(U_S^\times)$ and $w\in U_S$,
\be
(P_+x)(w)=\frac1{2\pi i}\oint_\gamma\tr^{(1)}x(z)^{(1)}r(z,w)dz.
\ee
\end{thm}
\begin{prf}
If $x\in\g(U_S)$, then the integrand has, by (i), only a pole at $w$ with residue
$\tr^{(1)}(x(w)^{(1)}C)=x(w)$. If $x$ is in the subspace spanned by the
$\xi_j$, the integral is zero by (iii). Finally, if 
$x(t)=\Ad( g(t)) y(t)$ with $y(t)$ holomorphic on $\Sigma-S$, then
\be
\oint_\gamma\tr^{(1)}x(t)^{(1)}r(t,w)\, dt=
\oint_\gamma\tr^{(1)}y(t)^{(1)}\Ad( g(t)^{-1})^{(1)}r(t,w)\,dt=0,
\ee
by Stokes' theorem,
since $\gamma$ is the boundary of a region 
where the integrand is a closed one-form.
\end{prf}

Let us now suppose that we have a local parametrization 
$\lambda\mapsto [P_{g_\lambda}]$ of the
moduli space of stable $G$-bundles by some region $\Lambda$ of $\C^m$:
for each $\lambda\in \Lambda$, we suppose to have a $g_\lambda\in
G(U_S^\times)$ holomorphic in $\lambda$, so that, for all
$\lambda\in \Lambda$  the $m$
vectors in $\g(U_S)$
\be
\xi_{\lambda j}=
\frac{\partial g_\lambda}{\partial\lambda_j}g^{-1}_\lambda
\ee
represent linearly independent classes in 
$H^{1}(\Sigma,\Ad(P_{g_\lambda}))$, the tangent space
 to the moduli space at $g_\lambda$.

Let us denote by $r(z,w;\lambda)dz$ the classical $r$-matrix
of Proposition \ref{prop12} with $g=g_\lambda$ and $\xi_j=\xi_{\lambda j}$.
Let $\omega_j(z,\lambda)dz$ denote
the dual basis in  $H^0(\Sigma,K\otimes\Ad(P_g))$.
Then we have the ``dynamical'' classical Yang--Baxter equation
\begin{thm}\label{cybe}
For all distinct $z_1,z_2,z_3\in U_S$,
\bea 
 &[r^{(13)}(z_1,z_3,\lambda),r^{(23)}(z_2,z_3,\lambda)]=& \\
 &[r^{(21)}(z_2,z_1,\lambda),r^{(13)}(z_1,z_3,\lambda)]
-[r^{(12)}(z_1,z_2,\lambda),r^{(23)}(z_2,z_3,\lambda)]& \\
&+\sum\omega_j^{(1)}(z_1,\lambda)\frac\partial{\partial\lambda_j}
r^{(23)}(z_2,z_3,\lambda)
-
\sum\omega_j^{(2)}(z_2,\lambda)\frac\partial{\partial\lambda_j}
r^{(13)}(z_1,z_3,\lambda).&
\eea
\end{thm}
The notation is that $t^{(ij)}=\sum x^{(i)}y^{(j)}$ if $t=\sum x\otimes y$.

\subsection{An explicit form for the connection}\label{tyty}
In this section we give an explicit form of the energy-momentum tensor
in terms of local coordinates on the moduli space of $G$-bundles.
It is given in terms of the dynamical classical $r$-matrix. 
Thus we consider, as before, a Riemann surface $\Sigma$, and an
open neighborhood
$U_S$ of a finite set $S\subset\Sigma$. Then for each $g\in G(U_S^\times)$
and each choice
of a set of vectors $\xi_1,\dots,\xi_m$ in $\g(U_S)^\times$ representing
a basis in $H^{1}(\Sigma, \Ad(P_g))$, we have an $r$-matrix $r(z,t)$.

The energy-momentum tensor is associated to a point  on 
some open set $U_0$, on which a local coordinate $z$ is chosen. 
To simplify the notation, we will simply write $z$, to denote a
point in $U_0$, which we view as a aubset of $\C$ via the local 
coordinate. We extend $g$ and $\xi_j$ to  maps
 from $U_{S}^\times\cup U_0$ by setting
$g=1$ and $\xi_j=0$ on $U_0$, and get in this way an $r$-matrix
$r(z,t)$ on $U_S\cup U_0$. By Lemma \ref{l99}, $r(z,t)dz$ for
$z,t\in U_0$ is the analytic continuation of
\be
\Ad(g(z)^{-1})\otimes\Ad(g(t)^{-1})(r(z,t)+\sum\omega_j(z)\otimes\xi_j(t))dz
\ee
from $U_S\times U_S$. We keep the notation $r(z,t)$ for this new
$r$-matrix since it coincides with the old one on $U_S\times U_S$.
We will also need $r(z,t)dz$ when $z\in U_0$ and $t\in U_S$. This
is the analytic continuation of $\Ad(g(z)^{-1})^{(1)}r(z,t)dz$.

Let $\lambda_1,\dots,\lambda_m$ be local complex coordinates on the moduli
space of stable $G$-bundles. We will fix for $\lambda\in\Lambda\subset\C^m$
an element $g_\lambda\in G(U^\times_S)$ holomorphically depending 
on $\lambda$ and
 representing the corresponding isomorphism
class of $G$-bundles. The image of the coordinate vector fields
$\partial/\partial \lambda_j$ are the classes in the tangent
space $H^1(\Sigma,\Ad(P_{g_\lambda}))$ at $[P_{g_\lambda}]$
represented by the vectors in $\g(U_S^\times)$
\be
\xi_{\lambda,j}(t)=\partial_{\lambda_j}g_\lambda(t)\,
g_\lambda(t)^{-1}.
\ee
 We also choose a holomorphic lift $\hat g_\lambda
=[(g_\lambda^{\mathrm{reg}},z_\lambda)]
\in G(U_S^\times)$, which amounts to choosing
a trivialization of the vector bundle of which conformal blocks are
sections. The choice of this lift is encoded in functions $a_j(\lambda)$
defined by 
\bea
D_{\xi_j}f(\hat g_\lambda)&=&-(\partial_{\lambda_j}
+ka_j(\lambda))f(\hat g_\lambda)\\
 &=_{\mathrm{def}}&-\nabla_{\lambda_j}f(\hat g_\lambda),
\eea
for any local holomorphic function $f$. Explicitly, 
\be
a_j(\lambda)=\frac1{2\pi i}\int_{U_S}
\tr(\partial(\partial_{\lambda_j}g_\lambda^{\mathrm{reg}}
g_\lambda^{{\mathrm{reg}}-1})\bar\partial 
g_\lambda^{\mathrm{reg}} g_\lambda^{{\mathrm{reg}}-1})
-\partial_{\lambda_j}\log(z_\lambda).
\ee
\vskip 30pt
\begin{definition} Let $r(z,t,\lambda)$ be the $r$-matrix corresponding to
 the data $g_\lambda$, $\xi_{\lambda,j}$.
Let
\be
r(z,t,\lambda)=\frac C{z-t}+r_0(z,\lambda)+O(t-z)
+(t-z)r_1(z,\lambda)+O((t-z)^2),
\ee
be the Laurent expansion of the classical $r$-matrix at its pole.

Let $q_1(z,\lambda)=[r_0(z,\lambda)]$, where $[\quad ]:\g\otimes\g
\to\g$ is the Lie bracket on $\g$.

Let $q_2(z,\lambda)=\mu(r_1(z,\lambda))$, where $\mu:\g\otimes\g
\to\C$ is the invariant symmetric bilinear form $x\otimes y\mapsto
\tr(xy)$.
\end{definition}

One essential ingredient is the $\ell$-operator. Its semiclassical
counterpart is a higher genus version of the Lax operator in
the $r$-matrix formulation of classical integrable systems.

\vskip 30pt
\begin{definition}
Let, for $z\in U_0$,
\begin{equation}\label{eqq3}
 q_3(z,\lambda)=\frac{-1}{2\pi i}\oint_\gamma\tr^{(2)}
 \bigl((r(z,t,\lambda)+\omega_j(z,\lambda)\otimes 
 \xi_{\lambda j}(t))(dg_\lambda(t)g_\lambda(t)^{-1})^{(2)}\bigr).
\end{equation}
The $\ell$-{\emph{operator}} is the differential operator in $\lambda$
 with values in $\g\otimes\End(V)$:
\be
\hat\ell(z)=
\sum_{j=1}^m\omega_j(z,\lambda)^{(0)}\nabla_{\lambda_j}
+kq_3(z,\lambda)^{(0)}+\sum_{j=1}^nr(z,p_j,\lambda)^{(0j)},
\ee
where the factors in $\g\otimes\End(V)=\g\otimes\End(V_1)\otimes
\cdots\otimes\End(V_n)$ are numbered from $0$ to $n$.
The ``spectral parameter'' $z$ runs over $U_0$.
\end{definition}
\begin{proposition}\label{key}
Let $z\in U_0$ and $\lambda\in \Lambda$. Set $\kappa=k+h^\vee$. Then
\begin{enumerate}
\item[(i)]
$J_x(z)u(\hat g_\lambda)=
-\tr^{(0)}(x^{(0)}\hat\ell(z))\,u(\hat g_\lambda)$
\item[(ii)] $
T(z)u(\hat g_\lambda)=
\frac1{2\kappa}\bigl(\tr^{(0)}\hat\ell(z)^2
+\tr^{(0)}q_1(z,\lambda)^{(0)}\hat\ell(z)
+k q_2(z,\lambda)\bigr)
u(\hat g_\lambda)
$
\end{enumerate}
\end{proposition}

\begin{prf}
We introduce the notation 
$\hat\ell_x(z)=\tr^{(0)}x^{(0)}\hat\ell(z)$.

(i)
By definition, $J_x(z)\,u(\lambda)=D_{x_{-1}(z)}i_z^{-1}u(i_z\hat
g_\lambda)$, where 
$x_{-1}(z)$ is the map $t\mapsto x/(t-z)$ for $t\in U_0$ and
is equal to zero on $U_S$.

To compute this, we have to decompose $x_{-1}(z)$ according to the 
decomposition \Ref{decomposition}: $x_{-1}(z)=y_++y_0+y_-$.

The first component is $y_+:=P_+x_{-1}(z)$. The contribution of this
piece to $J_x(z)u$ is $-\sum_{j=1}^ny_+(p_j)^{(j)}u(\hat g_\lambda)$.

By Theorem \ref{proj}, $y_+(p_j)$ is given by a contour integral over
$\gamma$. Since $x_+$ vanishes except on $U_0$, only the component 
$\gamma_z=\gamma\cap U_0$ contributes to the integral:
\be
y_+(p_j)=\frac1{2\pi i}
\oint_{\gamma_z}\tr^{(1)}\frac{x^{(1)}}{w-z}
\,r(w,p_j,\lambda)dw=\tr^{(1)}x\,r(z,p_j).\ee
The second component  is 
\be
y_0(t)=\sum_{j=1}^m\frac1{2\pi i}\oint_{\gamma_z}
\tr\left(\frac x{w-z}\,
{\omega_j(w)\,dw}\right)\,
\xi_{\lambda j}(t)=\sum_{j=1}^m\tr(x\,\omega_j(z))\xi_{\lambda j}(t).
\ee
Since $D_{\xi_{\lambda j}}i_z^{*-1}u\circ i_z=-\nabla_{\lambda_j}u$,
we get 
\be
D_{y_0}i_z^{*-1}u(i_z\hat g_\lambda)
=-\sum\tr(x\,\omega_j(z,\lambda))\partial_{\lambda_j}u(\hat g_\lambda).
\ee

We turn to the third component. By Lemma \ref{l4},
\begin{equation}\label{e962}
D_{y_-}i_z^{*-1}u(i_z\hat g_\lambda)
=-\frac k{2\pi i}\oint_{\gamma\cup\gamma_z}\tr(di_z(g_{\lambda})
i_z(g_\lambda)^{-1}y_-)u(\hat g_\lambda).
\end{equation}
Since $i_z(g_\lambda)$ is trivial on $U_0$, the integral
reduces to an integral over $\gamma$. By Theorem \ref{proj},
for $t\in U_S$,
\bea
y_-(t)&=&-y_0(t)-y_+(t)\\
 &=&-\sum_j\tr(\omega_j(z,\lambda)\,x)
\xi_{\lambda j}(t)-
\tr^{(1)}x^{(1)}r(z,t,\lambda).
\eea
Therefore,
\bea
D_{y_-}i_z^{*-1}u(i_z\hat g_\lambda)
&=&\frac1{2\pi i}\oint_{\gamma}\tr\otimes\tr
\bigl(x\otimes dg_{\lambda}(t)
g_\lambda(t)^{-1} \\
 & &(r(z,t,\lambda)+\omega_j(z,\lambda)\otimes
\xi_{\lambda j}(t))\bigr)u(\hat g_\lambda)
\\
 & &=-k\,\tr(x\,q_3(z,\lambda))u(\hat g_\lambda).
\eea
These terms taken together give $-\hat \ell$ after properly renumbering
the factors, as claimed.

\noi (ii) Let $x\in\g$, and $y=x_{-1}(z)$ (we will later take $x=b_j$ and
sum over $j$).
Thus
\be
y(t)=\left\{{\frac x{t-z},\qquad t\in U_0\\
          \atop           0, \qquad t\in U_S.}\right.
\ee
Let us decompose $y=y_++y_0+y_-$. Then
\be
D_y^2i_z^{*-1}u=D_{y_+}D_yi_z^{*-1}u+D_{y_0}D_yi_z^{*-1}u +
D_{y_-}D_yi_z^{*-1}u.
\ee
Let us consider the three terms one after the other.
Let us introduce the notation
 $\hat g_{\lambda,s}=\exp(-sy_+)i_z(\hat g_\lambda)$. 
We then have, by definition,
\begin{eqnarray}\label{e255}
D_{y_+}D_yi_z^{*-1}u(i_z\hat g_\lambda)&=&
\dds D_yi_z^{*-1}u(\hat g_{\lambda,s}) 
\nonumber
\\
 &=&\dds 
\bigl(-\hat\ell_x(z;\hat g_{\lambda,s},(\partial_{\lambda_j}g_{\lambda,s}
 g_{\lambda,s}^{-1}))i_z^{*-1}u(\hat g_{\lambda,s})
\\
 & &-\frac k{2\pi i}\oint_{\gamma_z}
\tr(dg_{\lambda_s}g_{\lambda,s}^{-1}y_-)u(\hat g_{\lambda})\bigr)
\nonumber\\
&=&A+B,\nonumber
\end{eqnarray}
where we have written the dependence of $\hat\ell_x(z)$ on the data
explicitly. The last term $B$ is the contribution of $\gamma_z$ to the
integral \Ref{e962}, which now does not vanish, since, on $U_0$,
$g_{\lambda,s}=\exp(-sy_+)\neq 1$. 
We can give a more explicit formula for this term:
\bea
B &=&-\dds\frac k{2\pi i}\oint_{\gamma_z}
\tr(dg_{\lambda_s}g_{\lambda,s}^{-1}y_-)\\
 &=&\frac k{2\pi i}\oint_{\gamma_z}\tr(dy_+y_-)\\
 &=&k\,\tr(y_+'(z)x),
\eea
by the residue theorem, since $y_-(t)=x/(t-z)+O(1)$, when $t\to z$.
Now,
\bea
y_+(t)&=&\frac x{t-z}+\tr^{(1)} x^{(1)}r(z,t,\lambda)\\
 &=&
\tr^{(1)} x^{(1)}r_0(z,\lambda)+(t-z)\tr^{(1)}
(x^{(1)}r_1(z,\lambda))+O((t-z)^2).
\eea
It follows that $y'_+(z)=\tr^{(1)}(x^{(1)}r_1(z,\lambda))$. Thus
\be
k\,\tr(y_+'(z)x)=k\,\tr\otimes\tr(x\otimes x\, r_1(z,\lambda)).
\ee
If $x=b_j$, the sum over $j$ of this expression is $k\mu(r_1(z,\lambda))$.

To deal with the first term in \Ref{e255}, we need to know how
$\hat \ell$ depends on $s$.
\begin{lemma} Let $g_\lambda\in \hat G(U_R^\times)$ 
  for all $\lambda\in\Lambda$ and $h\in G(U_R)$. Let $\tilde
  g_\lambda=hg_\lambda$, $\xi_{\lambda
    j}=\partial_{\lambda_j} g_\lambda g_\lambda^{-1}$ and
  $\tilde\xi_{\lambda j}=\partial_{\lambda_j} \tilde g_\lambda \tilde
  g_\lambda^{-1}$.  Then 
 \be \hat\ell_x(z;\tilde
  g_\lambda, (\tilde \xi_{\lambda j})) = \prod_{l=1}^n
  h(p_l)^{(l)}
  \hat\ell_{\Ad(h(z)^{-1})x}(z;g_\lambda,(\xi_{\lambda j}))
  \prod_{l=1}^n h(p_l)^{-1(l)}.  \ee
\end{lemma}
This follows from Lemma \ref{transf}, the following transformation
rule for the holomorphic one-forms $\omega_j(z;g_\lambda,(\xi_{\lambda j}))$:
\be
\omega_j(z;\tilde g_\lambda,(\tilde\xi_{\lambda j}))=
\Ad(h(z))\omega_j(z;g_\lambda,(\xi_{\lambda j})),
\ee
and the fact that if $\hat g_\lambda\to h\hat g_\lambda$,
\bea
a_j(\lambda)&\to& a_j(\lambda)+\frac1{2\pi i}
\oint_\gamma
\tr\bigl(\xi_{\lambda j}(t)h^{-1}dh(t)\bigr),
\\
q_3(z,\lambda)&\to& q_3(a,\lambda)+\sum_j\omega_j(z,\lambda)
\frac1{2\pi i}
\oint_\gamma
\tr\bigl(\xi_{\lambda j}(t)h^{-1}dh(t)\bigr),
\eea
leading to a cancellation.

In our case, $R=S\cup\{z\}$ and 
$h_=\exp(-sy_+)$. Using the invariance
$u(h g)=h\, u(g)$, we get
\bea
%D_{y_+}D_{y}i_z^{*-1}u(i_z\hat g_\lambda)&=&
A&=&-\dds \prod_{j=1}^n
\exp(-sy_+(p_j))^{(j)}
\hat\ell_{\Ad(\exp\,sy_+(z))x}(z)i_z^{*-1}u(\hat g_\lambda)
\\
 &=&
(-
\hat\ell_{[y_+(z),x]}(z)
+\sum_{l=1}^n
y_+(p_l)^{(l)}
)u(\hat g_\lambda)
\\
 &=&
-\tr^{(0)}\bigl(
[\tr^{(-1)}x^{(-1)}r_0(z,\lambda)^{(-1,0)},x^{(0)}]
\hat\ell(z)\bigr)
 u(g_\lambda)
\\
 & &+
\sum_j\tr^{(0)}\bigl(x^{(0)}r(z,t,\lambda)^{(0j)}\bigr)
\hat\ell_x(z)\,u(g_\lambda).
\eea
We have used here the fact that, for $s\in U_0$,
\bea
y_+(s)&=&\frac1{2\pi i}
\oint_{\gamma_z}\tr^{(1)}\frac{x^{(1)}}{t-z}r(t,s)\,dt \\
 &=&\frac x{s-z}+\tr^{(1)}x^{(1)}r(z,s).
\eea
In particular $y_+(z)=\tr^{(1)}x^{(1)}r_0(z,\lambda)$.
To avoid renumbering the factors, we gave the number ${-1}$ to the
first factor or $r_0$.
Note that, taking $x=b_j$, we have
\be
 \sum_{j}[\tr^{(-1)}b_j^{(-1)}r_0(z,\lambda)^{(-1,0)},b_j^{(0)}]
 =-[r(z,\lambda)]^{(0)}.
\ee
This follows from the identity $\tr(b_ja)[b,b_j]=[b,a]$, valid
for any $a\otimes b$.

The second term in the decomposition is treated easily:
\be
D_{y_0}D_yi_z^{*-1}u(i_z\hat g_\lambda)
=\sum_{j=1}^m\tr(x\,\omega_j(z,\lambda))\nabla_{\lambda_j}
(\hat\ell_x(z)u(g_\lambda))
\ee
We turn to the third term.
Since $D_{y}i_z^{*-1}u$ is invariant under the right action of
$G(\Sigma-(S\cup\{z\}))$, we have, by Lemma \ref{l4},
\be
D_{y_-}D_yi_z^{*-1}u(i_z\hat g_\lambda)=-\frac k{2\pi i}\oint_\gamma
\tr(dg_\lambda(t)g_\lambda(t)^{-1}y_-(t))D_yi_z^{*-1}u(i_z\hat g_\lambda).
\ee
The integral is actually over $\gamma\cup\gamma_z$, but the 
integral over $\gamma_z$ vanishes, since $g_\lambda$ is extended by
1 on $U_0^\times$. 
Since, for $t\in U_S$,
\be
y_-(t)=-\tr^{(1)}x^{(1)}r(z,t,\lambda)-\sum_j
\tr(x\,\omega_j(z,\lambda))\xi_{\lambda j}(t),
\ee
it follows that \be D_{y_-}D_yi_z^{*-1}u(i_z\hat g_\lambda)
=\tr(x\,q_3(z,\lambda))\hat\ell_x(z,\lambda)u(\hat g_\lambda).\ee
We then take all terms together, set $x=b_j$ and sum over $j$. 
\end{prf}

\subsection{Transformation properties}\label{herestheproof}

In this section we compute the dependence of the various
objects we introduced on the choices of coordinates and
trivializations. This computations will show that the 
connection is indeed well defined.

\begin{lemma}
The dependence of $q_1, q_2, q_3$ on the choice of coordinate
on $U_0$ is 
\bea
q_1^w(w,\lambda)&=&q_1^z(z,\lambda)\frac{dz}{dw}
\\
q_2^w(w,\lambda)&=&q_2^z(z,\lambda)\left(\frac{dz}{dw}\right)^2
+\frac{\dim(G)}6\{z,w\}
\\
q_3^w(w,\lambda)&=&q_3^z(z,\lambda)\frac{dz}{dw}
\eea
\end{lemma}
To prove this lemma one uses that $r(z,t)$ depends on coordinates
 locally as a  one-form
in the first argument and as a function in the second. Then one
compares the Laurent expansions of $r$ in two different coordinates.
This gives the formulae for $q_1, q_2$. As for $q_3$, the statement
is obvious.

Let us from now on assume that all coordinates we consider are
part of a flat atlas, so that Schwarzian derivatives do not appear.
Then $q_1dz$, $q_3dz$ have an analytic continuation to holomorphic
1-forms away from $S$, and $q_2dz^2$ has an analytic continuation
to a holomorphic quadratic differential away form $S$. The
singularity at $S$ can be arbitrary. However, using the 
transformation behavior of $r$, we deduce

\begin{proposition}\label{pqj}
$q_1dz$, $q_2dz^2$, $q_3dz$ extend to holomorphic (quadratic) 
differentials on $\Sigma-S$. Moreover, for $z\in U_S^\times$, and if
we denote the derivative with respect to $z$ by a prime,
\bea
 &\Ad(g_\lambda(z))q_1(z,\lambda)dz
+2 h^\vee dg_\lambda(z)g_\lambda(z)^{-1}
-
\sum [\omega_j(z,\lambda),\xi_{\lambda j}(z)]dz,&
\\
 &\bigl(q_2(z,\lambda)
-\sum\tr(\omega_j(z,\lambda)\xi'_j(z,\lambda))
+\tr g_\lambda(z)^{-1}g'_\lambda(z)q_1(z,\lambda)& \\
 &
-h^\vee\tr\bigl((g_\lambda'(z)g_\lambda(z)^{-1})^2\bigr)\bigr)dz^2,&
\\
 & \Ad(g_\lambda(z))q_3(z,\lambda)dz +dg_\lambda(z)g_\lambda(z)^{-1},&
\eea
extend to holomorphic (quadratic) differentials $q^S_{1}dz$,
$q^S_{2}dz^2$, $q^S_{3}dz$ on $U_S$.
\end{proposition}
\noi{\bf Remark.}\ $q^S_{1}=[r_0]$ and $q^S_{2}=\mu(r_1)$ 
where $r_0$ and $r_1$ are the Laurent coefficients of the
$r$-matrix on $U_S$. Also, $q^S_{3}$ is given by the integral formula
\Ref{eqq3} but with an integration contour that encircles $z$, as
well as the points $p_j$.

 Finally, we need the transformation behavior of $\hat\ell$ 
which can be readily deduced from the one of $r$ and $q_3$.

\begin{proposition}
$\hat\ell(z)dz$ has an analytic continuation as a function
of $z$ to a one-form on $\Sigma-S$. Moreover, for $z\in U^\times_S$,
\be
\hat\ell_S(z)=\Ad(g_\lambda(z))^{(0)}
\hat\ell(z)dz+kdg_\lambda(z)g_\lambda(z)^{-1(0)}
\ee
extends to a meromorphic one-form on $U_S$ with at most simple poles
on $S$.
\end{proposition}

These results prove in particular part (i) of Proposition 
\ref{wherestheproof}.
Part (ii) follows by computing the left derivative (or the
right derivative) along the lines of the proof of Proposition
\ref{key}
(i), i.e., by decomposing $y$ into the three components and
using the invariance properties of conformal block to express
the action in terms of $\ell$-operators.

 Let us put everything together. We have the second order differential
operator $A(z)$ in the variables $\lambda_j$ acting on $V$-valued functions,
and defined a priori for $z\in U_0$:
\be
A(z)=\frac1{2\kappa}\bigl(\tr^{(0)}\hat\ell(z)^2+
\tr^{(0)}q_1(z,\lambda)^{(0)}\hat\ell(z)+kq_2(z,\lambda)\bigr)
\ee
\begin{corollary}
$A(z)dz^2$ extends (given a flat structure) to a quadratic 
differential on $\Sigma-S$. Moreover, if $z\in U_S^\times$,
\be
A_S(z)=A(z)+\tr^{(0)}((g'_\lambda(z)g_\lambda(z)^{-1})^{(0)}\hat\ell(z))
+\frac k2\tr((g_\lambda'(z)g_\lambda(z)^{-1})^2)
\ee
extends to a meromorphic quadratic differential on $U_S$ with
at most poles of second order on $S$.
\end{corollary}
In particular, this proves Proposition \ref{wheresitsproof}.

\section{Moving points}

We describe here the KZB equations that correspond to moving
the points and the tangent vectors, but keeping the complex
structure on $\Sigma$ fixed. 

\subsection{Fixing the complex structure} 
Let us thus fix a Riemann surface
$(\Sigma,I)$ of genus $h$ with complex structure $I$ and let 
$B_n$ be the moduli space of data $(q_i,w_i)_{i=1}^n$
consisting of $n$ distinct points $q_i$ on $\Sigma$  and 
$n$ non-zero tangent vectors $w_i\in T_{q_i}\Sigma$, modulo
the natural action of the group of conformal automorphisms of $\Sigma$.
If $n\geq 1$ ($n\geq 2$ for genus zero), $B_n$ is the smooth
algebraic variety
\be
B_n=T^\times(\Sigma^n-\cup_{i<j}\{x_i=x_j\})/{\mathrm{Aut}}(\Sigma),
\ee
where $T^\times$ denotes the complement  in the holomorphic tangent bundle
of the  set of vectors with
at least one vanishing component.

We have an embbedding $j:B_n\hookrightarrow \hat M_{h,n}$ sending
the class of $(q_i,w_i)$ to the class of $\phi_*(I)$ for any
diffeomorphism $\phi:\Sigma\to\Sigma$ such that  $\phi(q_i)=p_i$ and
$\phi_*(q_i)w_i=v_i$ for all $i=1,\dots,n$.

Therefore the connection on conformal blocks  induces a connection
on the pull-back of the projectivized vector bundle of conformal
blocks on $B_n$. To describe this connection, we need to study
the tangent map $j_*$ of $j$. Let $(\dot q_i,\dot w_i)$
be a tangent vector at $(q_i,w_i)$ and let $(q_i(s),w_i(s))$, $|s|<\epsilon$,
represent a curve  in $B_n$ with tangent vector $(\dot q_i,\dot w_i)$ at
$s=0$. Suppose $\phi$ is a diffeomorphism
sending $(q_i,w_i)$ to $(p_i,v_i)$, and let $z:U_S\to \sqcup_{j=1}^n\C$ 
be a local complex coordinate for $\phi_*(I)$ on some neighborhood
$U_S$ of $S$ as above.          This coordinate pulls back to
a local flat 
coordinate $z\circ \phi$ on a neighborhood of $\{q_1,\dots,q_n\}$.
Let $z_i(s)=z\circ\phi(q_i(s))$, $\eta_i(s)
\frac d{dz}=z\circ\phi_*(q_i(s))w_i$
be the expression of our family with respect to this local coordinate.
By construction, $z_i(0)=0$, $\eta_i(0)=1$. For $|s|<\epsilon$,
let us choose a diffeomorphism $\phi_s$ of $\Sigma$ such that 
$z(\phi_s(q))=\eta_i(s)^{-1}(z(\phi(q))-z_i(s))$ for $q$  close
to $q_i$, and such that $\phi_0=\phi$. Such diffeomorphisms can be
easily constructed by taking $\phi_s=\phi$ except on some neighborhood
of the points $q_i$ and choosing a suitable interpolation in an
annular region around the points. The diffeomorphism $\phi_s$ 
sends $(q_i(s),w_i(s))$ to $(p_i,v_i)$, so it can be used
to define $j(q_i(s),w_i(s))={\mathrm{cls}}(\phi_{s*}(I))$.
Then $z$ is still a local complex coordinate in a sufficiently
small neighborhood of $S$ for all complex structures $\phi_{s*}(I))$,
and we have a curve $(\phi_{s*}(I),z)$ in the enlarged moduli
space (see \ref{cobops}).

The tangent vector $\dot I$ to the curve  $\phi_{s*}(I)$ corresponds
to a Beltrami differential $\mu=-\bar\partial\xi$  where $\xi$ is
the vector field on $\Sigma$ such that $\xi(\phi(x))=
\frac d{ds}|_{s=0}\phi_s(x)$. This vector field is holomorphic
on a neighborhood $U_S$ of $S$, thus $\mu=0$ on $U_S$. The
tangent vector $(\dot I,\dot z=0)$ to the curve in the enlarged
moduli space is then represented by the holomorphic vector field 
$\zeta=\xi$ on $U_S^\times$. Indeed, we can write $\mu=\bar\partial
\zeta_S$ on $U_S$ and $\mu=\bar\partial\zeta_\infty$ on the complement of 
$S$ with $\zeta_S=0$ and $\zeta_\infty=-\xi$. Thus $\zeta=\zeta_S-\zeta_\infty
=\xi$.

Let us summarize.

\begin{proposition}
Let us fix some coordinate $z:U_S\to\sqcup_{j=1}^n\C$ on some neighborhood of
$S$ and let $z_1,\dots,z_n,\eta_1,\dots,\eta_n$ be the coordinates of
 a point in $B_n$ in  neighborhood of $(p_i,v_i)$.
Then the covariant derivative in the direction of a tangent vector
$(\dot z_i,\dot \eta_i)$ is given by
\bea
\nabla_{\dot z_i,\dot \eta_i}u(z_i,\eta_i,\hat g_\lambda)&=&
\sum \dot z_i\frac{\partial}{\partial z_i}u(z_i,\eta_i,\hat g_\lambda)\\
 & &+\sum \dot \eta_i
\frac{\partial}{\partial{\eta_i}}u(z_i,\eta_i,\hat g_\lambda)
\\
 & &+\frac1{2\pi i}
\oint_\gamma T^S(z)\zeta(z)dz\,u(z_i,\eta_i,\hat g_\lambda),
\eea
where $\zeta=-(\dot\eta_iz+\dot z_i)\ddz$
\end{proposition}

The fact that $T^S$ rather than $T$ appears in this formula is due 
to the choice of describing the bundle by a function $g_\lambda$ 
which is a fixed function of the coordinate $z$
when we move the points, whereas in \Ref{e300} $g_\lambda$ is
fixed in the coordinate vanishing at $S$. The difference is given
in terms of $J$ and cancels the terms in Proposition \ref{wheresitsproof}.
The details are left to the reader.

We now give a more explicit formula for the connection in this
case. We keep the notation of the previous proposition, and
express our connection in terms of the $r$-matrix $r(z,w,\lambda)$,
$z,w\in U_S$, its constant term $r_0$, defined by
\be
r(z,w,\lambda)=\frac C{z-w}+r_0(z,\lambda)+O(w-z),
\ee
and the one-form $q(z,\lambda)=\sum\omega_\nu(z,\lambda)a_\nu(\lambda)
+q^S_3(z,\lambda)$, where $a_\nu(\lambda)$ depend on the choice
of local trivialization of the $\C^\times$-bundle $\hat G$, see
\ref{tyty}, and $q_3^S$ may be characterized by the properties:
\begin{enumerate}
\item[(i)] $q_3^S(z,\lambda)$ is holomorphic on $U_S$.
\item[(ii)]
$\Ad(g_\lambda(z)^{-1})q_3^S(z,\lambda)dz-g_\lambda(z)^{-1}dg_\lambda(z)$
extends to a holomorphic one-form on $\Sigma-S$.
\item[(iii)]$\oint_\gamma\tr(q_3^S\xi_\nu)dz=0$ for all $\nu=1,\dots, m$.
\end{enumerate}
Note that $q_3^S$ can also be expressed in terms of the $r$-matrix,
and $g_\lambda$, 
see the remark after Proposition \ref{pqj}.

We will denote by ${\mathrm{Cas}}(V_j)$ the value of the central
Casimir element $\sum b_i^2$ on the irreducible representation 
$V_i$.
Evaluating the contour integrals, we get

\begin{thm}\label{tkz}
The connection on the space of conformal blocks
restricted to
$B_n$ is given by the formula:
\be
\nabla=\sum dz_i\nabla_{z_i}+\sum d\eta_i\nabla_{\eta_i}
\ee
where
\bea
\nabla_{z_i}
 &=&\frac\partial{\partial z_i}
-\frac1\kappa\biggl(\sum_{\nu=1}^n\omega_\nu(z_i,\lambda)^{(i)}
\frac{\partial}{\partial\lambda_\nu}
+
k q(z_i,\lambda)^{(i)}
\\
 & &+r_0(z_i,\lambda)^{(ii)}
+\sum_{j:j\neq i}r(z_i,z_j,\lambda)^{(ij)}\biggr),
\eea
and 
\be
\nabla_{\eta_i}=
\frac{\partial}{\partial\eta_i}
-\frac{{\mathrm{Cas}}(V_i)}{2\kappa}
\ee
\end{thm}

The next result is that this connection is flat. In fact, more
strongly, we have the following result for the differential
operators $\nabla_{z_i}, \nabla_{\eta_i}$ acting on any functions
of $z,\eta,\lambda$, not just conformal blocks.

\begin{thm}
The differential operators $\nabla_{z_i},\nabla_{\eta_i}$ with
coefficients in $U(\g)^{\otimes n}$ defined
by the formulae in Theorem \ref{tkz} commute with each other for any 
complex value of $k\neq -h^\vee$.
\end{thm}

In particular, we may take the limit $k\to -h^\vee$ and obtain, for
each $(z_i,\eta_i)$, a set of commuting differential operators
in the coordinates $\lambda$, and
forming a generalization of the integrable Gaudin model:

\begin{corollary}
For each $z_1,\dots, z_n$, the differential operators
\be
H_i=
\sum_{\nu=1}^n\omega_\nu(z_i,\lambda)^{(i)}
\frac{\partial}{\partial\lambda_\nu}
-
h^\vee q(z_i,\lambda)^{(i)}
+r_0(z_i,\lambda)^{(ii)}
+\sum_{j:j\neq i}r(z_i,z_j,\lambda)^{(ij)},
\ee
commute with each other.
\end{corollary}

\subsection{Proof of Theorem \ref{tkz}}
The only thing that we have to prove is that 
$\nabla_{z_j}\nabla_{z_l}=
\nabla_{z_l}\nabla_{z_j}$.
We first remark that this follows from the universal
(i.e., independent of the representations $V_i$) identities:
\begin{eqnarray}
 &\displaystyle\frac{\partial A_1}{\partial z_2}-
\frac{\partial A_2}{\partial z_1}
=0,&\label{uno}
\\
 &\displaystyle
\kappa\left(\frac{\partial r^{(12)}}{\partial z_2}
-
\frac{\partial r^{(21)}}{\partial z_1}\right)
+\label{due}
[A_1+r^{(12)},A_2+r^{(21)}]=0,
& \\ &
[A_1,r^{(23)}]+
[r^{(13)},A_2]+\label{tre}
[r^{(13)},r^{(23)}]& 
\\
 &+
[r^{(12)},r^{(23)}]-
[r^{(21)},r^{(13)}]=0. & \nonumber
\end{eqnarray}
Here we have used the abbreviations
$r^{(ij)}=r(z_i,z_j,\lambda)^{(ij)}\in \g\otimes\g$ and
\be
A_j=\sum\omega_\nu(z_j,\lambda)^{(j)}
\frac{\partial}{\partial \lambda_\nu}+k\,q(z_j,\lambda)^{(j)}+
r_0(z_j,\lambda)^{(jj)}.
\ee
The differential operators $A_j$ have coefficients in the 
universal enveloping algebra of $\g$.

In the last of these identities we recognize the dynamical
classical Yang--Baxter equation, Theorem \ref{cybe}. The
first identity follows trivially from the fact that 
$A_i$ is independent of $z_j$, if $i\neq j$. What is
left to prove is the identity  \Ref{due}.
The left hand side of this identity consists of three
parts that, as we now show, vanish separately: the
first part is the homogeneous first order part of this
first order differential operator. The second part
is of zero order as a differential operator and is
proportional to $k$. The third part is of zero order
and independent of $k$.

The vanishing of these three parts is the respective contents of
the next three lemmata. We write the formulae in an abbreviated 
notation: thus $r^{(ij)}$, $q^{(i)}$, $r_0^{(ii)}$, $\omega_\nu^{(i)}$
stand for $r^{(ij)}(z_i,z_j,\lambda)$, $q^{(i)}(z_i,\lambda)$,
and so on.

\begin{lemma} For all $\mu=1,\dots,m$,
\be
\sum_\nu\omega_\nu^{(1)}\frac{\partial\omega_\mu^{(2)}}
{\partial\lambda_\nu}
-
\sum_\nu
\frac{\partial\omega_\mu^{(1)}}
{\partial\lambda_\nu}
\omega_\nu^{(2)}
+[\omega_\mu^{(1)},r^{(21)}]
+[r^{(12)},\omega_\mu^{(2)}]=0.
\ee
\end{lemma}
\begin{prf}
The left-hand  side is regular at $z_1=z_2$:
the poles cancel by the invariance property
$[C,x^{(1)}+x^{(2)}]=0$ valid for any $x\in\g$.
 Let us consider this left-hand side as a function of
$z_1$. We use the properties of the $r$-matrices: we know
that $\Ad(g(z_1)^{-1})^{(1)}r^{(12)}dz_1$ extends to
a holomorphic one-form on $\Sigma-S$ and that $\Ad(g(z_1)^{-1})^{(1)}
(r^{(21)}+\sum\xi_\nu^{(1)}\omega_\nu^{(2)})$ extends
to a holomorphic function on $\Sigma-S$.  Also,
$\Ad(g(z_1)^{-1})^{(1)}\omega^{(1)}dz_1$ extends to a 
$\g$-valued holomorphic one-form on $\Sigma-S$. By putting these
things
together, we see that acting with $\Ad(g(z_1)^{-1})^{(1)}$  on
the left-hand side of the identity   yields a holomorphic
one-form in $z_1$ on $\Sigma-S$. A similar conclusion holds
for $z_2$ and the second factor. Hence the left-hand side
has the form
\begin{equation}\label{e707}
\sum_{\mu,\nu}B_{\mu,\nu}
\omega_\mu^{(1)}
\omega_\nu^{(2)}.
\end{equation}
The constant coefficients $B_{\mu,\nu}$ can be calculated as
contour integrals over $\gamma\times\gamma$
of the trace of the left-hand side times $\xi_\mu^{(1)}\xi_\nu^{(2)}$.
The latter integral can be evaluated using the defining property  (iii) 
of the $r$-matrix and the definition of the one-forms $\omega_\nu$.
The result is that the integral vanishes for all $\mu,\nu$. Hence
\Ref{e707} is zero and the claim follows.
\end{prf}

\begin{lemma}
\be
\frac{\partial r^{(12)}}{\partial z_2}
-
\frac{\partial r^{(21)}}{\partial z_1} 
+
\sum_{\nu}\biggl(\omega_\nu^{(1)}
\frac{\partial q^{(2)}}{\partial\lambda_\nu}
-
\frac{\partial q^{(1)}}{\partial\lambda_\nu}
\omega_\nu^{(2)} \biggr)
+[q^{(1)},r^{(21)}]
+[r^{(12)},q^{(2)}]=0.
\ee
\end{lemma}

\begin{prf} This lemma is proved in a similar way as the previous
one: one uses the transformation and analytic continuation properties
of the various objects, to show that the left-hand side is of
the form \Ref{e707} and then shows that the coefficients
$B_{\mu,\nu}$ vanish, by computing them as contour integrals.
The calculation reduces to
\be
B_{\mu,\nu}=
\frac{\partial a_\mu}
{\partial \lambda_\nu}
-
\frac{\partial a_\nu}
{\partial \lambda_\mu}
-
\frac1{2\pi i}\oint_\gamma\tr(d\xi_\mu\xi_\nu).
\ee
But this is zero by Proposition \ref{propLie} and the definition
of $a_\nu$ in terms of $D_{\xi_\nu}$.
\end{prf}

\begin{lemma}
\bea
 &\displaystyle
h^\vee\left(\frac{\partial r^{(12)}}{\partial z_2}
-
\frac{\partial r^{(21)}}{\partial z_1}\right)
+[r_0^{(11)},r^{(21)}]
+[r^{(12)},r_0^{(22)}]
+[r^{(12)},r^{(21)}]
& 
 \\
 &
+
\sum_{\nu}\left(
\omega_\nu^{(1)}\frac{\partial}{\partial\lambda_\nu}
(r_0^{(22)}
+ r^{(21)})
-
\omega_\nu^{(2)}\frac{\partial}{\partial\lambda_\nu}
(r_0^{(11)}
+ r^{(12)})\right)
=0.& 
\eea
\end{lemma}

\begin{prf}
This is a degenerate case of the dynamical classical Yang--Baxter
equation. Let $m_{13}$, $m_{23}\in{\mathrm{Hom}}(U\g^{\otimes 3},
U\g^{\otimes 2})$ be the linear map on the third tensor power of
the universal enveloping algebra such that 
\be
m_{13}(x\otimes y\otimes z)=xz\otimes y,\qquad
m_{23}(x\otimes y\otimes z)=x\otimes yz.
\ee
If $t(z_1,z_2,z_3)\in\g^{\otimes 3}\subset U\g^{\otimes 3}$ denotes
the difference between the two sides of the dynamical classical
Yang--Baxter equation, then the identity claimed in the lemma
is equivalent to
\be
m_{13}t(z_1,z_2,z_1)+m_{23}t(z_1,z_2,z_2)=0,
\ee
and thus follows from Theorem \ref{cybe}.
The straightforward details are left to the reader. Here we
will only explain the appearance of the dual Coxeter number
in the formula. One of the terms that appear by taking
the limit $z_3\to z_1$ in the Yang--Baxter tensor is
\be
-m_{13}[C^{(13)},\partial_{z_3}r^{(23)}]=
-\sum b_i[b_i,y_j]\otimes x_j,\qquad{\mathrm{if}}\;
\partial_{z_3}r(z_2,z_3,\lambda)=\sum x_j\otimes y_j.
\ee
The dual Coxeter number is half the value of the
Casimir element in the adjoint representation. It appears
in this formula since, for any $x\in\g$, we have the identity
in the universal enveloping algebra
\bea
\sum b_i[b_i,x]&=&{\textstyle\frac12}\sum
\left([b_i^2,x]+[b_i,[b_i,x]]\right)
\\
&=&{\textstyle\frac12}(0+2\,h^\vee x)
\\
&=&h^\vee x.
\eea
\end{prf}

\section{Concluding remarks}
In this paper we have written the Knizhnik--Zamolodchikov--Bernard
equations in arbitrary local coordinates, and checked the
independence of choices. 

 We have  addressed the question of flatness of the connection
only in the situation where we move the marked points by keeping
the complex structure fixed.
 In fact this connection is supposed to be flat (on the
projectivization of the vector bundle of conformal blocks) on
the whole moduli space.
In the case of no marked points, Hitchin's proof \cite{H1}
of the flatness goes as follows. The principal symbol
of the covariant derivatives $\nabla_j$ in the direction
of some coordinate basis of the tangent space are Poisson commuting.
In fact they are  commuting Hamiltonians of Hitchin
systems \cite{H2}. This property implies that the
curvature of the connection is in fact a differential
operator of second order rather than of third order, as one
would a priori expect. Then one uses cohomological arguments
to show that such that there are in fact no globally defined
second order operators except for constants.

In our situation we also have the Poisson commutation
of a principal part of the operators, as can be shown by
$r$-martix techniques. In fact they describe the
second order integrals of motion of Hitchin systems for
curves with marked points. However the cohomological 
arguments are more tricky in this case.

A direct general proof, based on the classical Yang--Baxter equation,
Theorem \ref{cybe}, would be instructive, and would
have the advantage to give commutation of the differential
operators $\nabla_j$ for general complex values of $\kappa$,
and when acting on more general functions than conformal blocks,
for which the differential operators can be considered.

In fact, Theorem \ref{tkz} gives flatness in the directions
of fixed complex structure in a universal form: it expresses the
commutativity of the covariant derivatives, viewed as differential
operators with coefficients in a tensor power of the universal 
enveloping algebra, without the need to mention representations.

In particular one could approach the interesting point
$\kappa=0$, which is related to a quantization of
Hitchin systems, see \cite{H1,H2} and to the Beilinson--Drinfeld
geometric Langlands correspondence, see \cite{Fr}.

It is likely that the proof of Theorem \ref{tkz} can be applied
to prove the flatness of the connection on the whole moduli space,
but the technical details appear to be more involved.

One motivation for the construction described in this paper is
the hope to understand the $q$-deformation of conformal field
theory on Riemann surfaces. It turns out that in special cases
the KZB equations admit a $q$-deformation, a system of 
compatible {\emph{difference}} equations: see \cite{FR} and Varchenko's 
contribution to these proceedings for the genus zero case, and
\cite{F1,FVT}, for the genus one case. In this paper, we have
completed the first step in the ``St.~Petersburg
$q$-deformation recipe'' (see \cite{FT} and Faddeev's lectures
in these proceedings): we have written all equations in terms
of (a version of) classical $r$-matrices. The second step is
to replace the classical $r$-matrices by quantum $R$-matrices,
which for genera larger than one is an open problem.

\vskip 40pt
\noindent{\it Acknowledgments.}
 I wish to thank K. Gaw\c edzki and A. Varchenko
for many useful discussions on this and related subjects, 
B. Enriquez for interesting correspondence and S. Kumar
for his explanations.
Also I am grateful to the organizers and the participants to the les Houches
school for the pleasant and stimulating atmosphere.
This research was partially supported by the NSF
grant DMS-9400841.

\end{document}